\documentclass[aps,prl,amsmath,amssymb,twocolumn,groupedaddress,longbibliography,floats,final,postprint,superscriptaddress]{revtex4-1}
\usepackage{hyperref}
\usepackage{graphicx}
\usepackage[utf8]{inputenc}
\usepackage{amsmath}
\usepackage{mathtools}
\usepackage{algorithm}
\usepackage[noend]{algpseudocode}
\usepackage{verbatim}
\usepackage{amssymb}
\usepackage{color}
\usepackage{mathtools}
\usepackage{bm}
\usepackage{stmaryrd}
\usepackage{subfigure}
\usepackage{tikz}
\usepackage{ifthen}
\hypersetup{
    colorlinks=true,
    linkcolor=blue,
    filecolor=gray,
    urlcolor=blue,
    citecolor=blue,
    }
\usetikzlibrary{calc}

\def\intv#1[#2..#3]{\llbracket #2\mathrel{{.}\,{.}}\nobreak#3\rrbracket}

\begin{document}

\date{\today}\title{Loop-Cluster Coupling and Algorithm for Classical Statistical Models}
\author{Lei Zhang}
\affiliation{Hefei National Laboratory for Physical Sciences at Microscale and Department of Modern Physics, University of Science and Technology of China, Hefei, Anhui 230026, China}
\affiliation{CAS Center for Excellence and Synergetic Innovation Center in Quantum Information and Quantum Physics, University of Science and Technology of China, Hefei, Anhui 230026, China}
\author{Manon Michel}
\email{manon.michel@uca.fr}
\affiliation{CNRS, Laboratoire de math\'ematiques Blaise Pascal, UMR 6620, Universit\'e Clermont-Auvergne, Aubi\`ere, France}
\author{Eren M. El{\c{c}}i}
\email{elci@posteo.de}
\affiliation{School of Mathematical Sciences, Monash University, Clayton, VIC 3800, Australia}
\author{Youjin Deng}
\email{yjdeng@ustc.edu.cn}
\affiliation{Hefei National Laboratory for Physical Sciences at Microscale and Department of Modern Physics, University of Science and Technology of China, Hefei, Anhui 230026, China}
\affiliation{CAS Center for Excellence and Synergetic Innovation Center in Quantum Information and Quantum Physics, University of Science and Technology of China, Hefei, Anhui 230026, China}
\affiliation{Department of Physics and Electronic Information Engineering, Minjiang University, Fuzhou, Fujian 350108, China}

\begin{abstract}
  Potts spin systems play a fundamental role in statistical mechanics
  and quantum field theory, and can be studied within the spin, the
  Fortuin-Kasteleyn (FK) bond or the $q$-flow (loop) representation.
  We introduce a Loop-Cluster (LC) joint model of bond-occupation
  variables interacting with $q$-flow variables, and formulate a LC
  algorithm that is found to be in the same dynamical universality as
  the celebrated Swendsen-Wang algorithm.  This leads to a theoretical
  unification for all the representations, and numerically, one can
  apply the most efficient algorithm in one representation and measure
  physical quantities in others.  Moreover, by using the LC scheme, we
  construct a hierarchy of geometric objects that contain as special
  cases the $q$-flow clusters and the backbone of FK clusters,
  the exact values of whose fractal dimensions in two dimensions
  remain as an open question.  Our work not only provides a unified
  framework and an efficient algorithm for the Potts model, but also
  brings new insights into rich geometric structures of the FK clusters.
\end{abstract}
%\pacs{%02.70.Tt, 05.10.Ln, 05.10.-a, 64.60.De, 75.10.Hk, 75.10.Nr
%}

%\keywords{%Monte Carlo methods;
%}

\maketitle

{\it Introduction.} The understanding of critical phenomena is now
strongly intertwined with the study of the rich behavior of the
$q$-state Potts model \cite{Wu_1982}.  Aside from the historical spin
representation~\cite{baxter2016exactly,nienhuis1984critical},
two other representations of the Potts model have played a central role:
the $q$-flow representation~\cite{essam1986potts,wu1988potts},
  which is a generalization of the loop description, and the Fortuin-Kasteleyn
  (FK) bond representation~\cite{kasteleyn1969phase,fortuin1972random}, which is also
  known as the random-cluster (RC) model.  On one hand, theoretical
advances were achieved thanks to the geometric and probabilistic
interpretations they brought, as well as the extension to positive
real $q$ values~\cite{chayes1998graphical,deng2007cluster,deng2007critical}.
For instance, they play an important role in conformal field
theory~\cite{di1997conformal} and in stochastic Loewner
evolution~\cite{schramm2000schramm,rohde2005s,lawler2005conformally,kager2004guide,cardy2005sle}.
On the other hand, numerical Monte Carlo (MC) methods, decisive in the
study of not-exactly soluble models, have significantly
 benefitted from these insights. Indeed,
the Metropolis \cite{metropolis1953equation} or heat-bath schemes rely
on single-spin moves and often suffer from severe \emph{critical
slowing-down}~\cite{hohenberg1977theory,sokal1997functional},
and the Sweeny algorithm~\cite{sweeny1983monte}, a local-bond update scheme,
has complications from connectivity-checking.
Based on the coupling between spin and FK representations~\cite{kasteleyn1969phase,fortuin1972random,edwards1988generalization},
efficient cluster methods,
including the  Swendsen-Wang (SW) and Wolff algorithms~\cite{swendsen1987nonuniversal,wolff1989collective},
have been developed and widely used. For the $q$-flow representation, one can apply the Prokof’ev-Svistunov worm algorithm~\cite{prokof2001worm,prokof1998worm,mercado2012worm,elcci2018lifted},
  which has proven to be particularly efficient at
  computing the magnetic susceptibility~\cite{deng2007dynamic} and the
  spin-spin correlation function~\cite{wolff2009simulating}.

However, despite the existence of the coupling between spin and FK
  representation for decades~\cite{kasteleyn1969phase,fortuin1972random,edwards1988generalization},
  a generic coupling between the $q$-flow and another representation,
  which would tie the three representations of the Potts model together,
  has remained an open question.

\begin{figure}[!ht]
  \includegraphics[width=1.0\columnwidth]{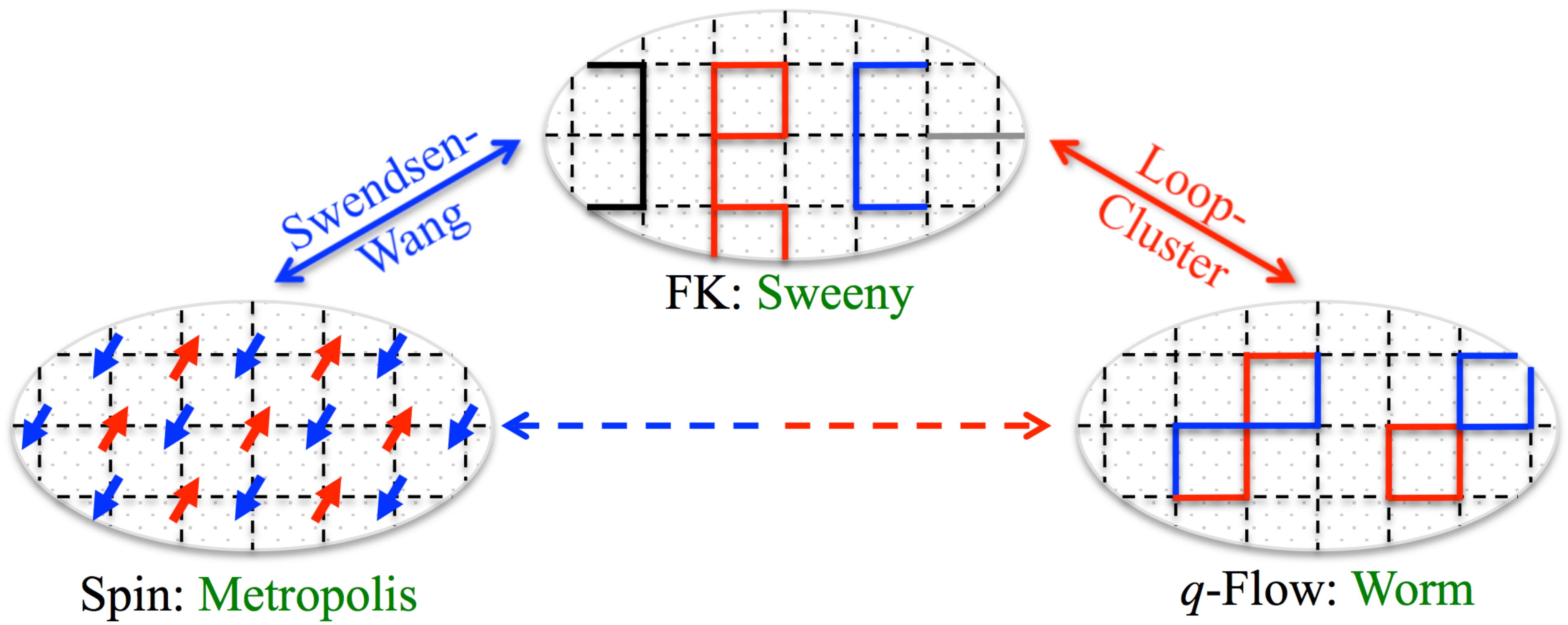}
  \caption{Representations and algorithms for the Potts model.  The
    spin, $q$-flow, and FK representations are coupled by the
    combination of the Swendsen-Wang and the Loop-Cluster algorithm.}
  \vspace{-5mm}
  \label{fig:tri}
\end{figure}

  In this Letter, we propose a unified framework by introducing a
  joint model, called the Loop-Cluster (LC) model, of FK bond
  variables interacting with $q$-flow variables. It includes and
  provides a straightforward derivation of the coupling for the Ising
  model~\cite{grimmett2009random,Evertz_2002},
  and applies to the Potts model of any integer $q \! \geq \!  1$.
  The LC joint model provides a setup for a new MC algorithm, which we
  call the Loop-Cluster (LC) algorithm.
  By investigating the dynamical properties over the complete graph and
  $d=2,3,4,5$ toroidals grids, we show that the LC and the SW schemes
  are in the same universality class.
  As a consequence, the three representations are tied together, and
  numerically, one can apply the most efficient algorithm in one
  representation and measure observables in others, as illustrated in Fig.~\ref{fig:tri}.

  Much insight is also gained on geometric structures of the Potts model from the LC scheme.
  The $q$-flow clusters, defined by sets of
  vertices connected by non-zero flow variables, can be proven to be
  contained in the backbones of FK clusters.
  Further,  we construct a hierarchy of random
  $q_{\rm F}$-flow clusters from a $q$-state FK configuration with real $q \geq 0$
  and integer $q_{\rm F} \geq 2$,
  which reduce to the $q$-flow clusters for $q_{\rm F}=q$
  and the backbones for $q_{\rm F} \! \rightarrow \! \infty$.
  This provides a new perspective to study the long-standing question about the backbone dimension
  for percolation and FK  clusters~\cite{Grassberger_1999,Smirnov_2001,Jacobsen_2002,Jacobsen_2002_2,Deng_2004,Xu_2014,Elci_2016}.
  In two dimensions (2D), We determine with high precision the fractal dimension $d_{\rm F}$ for various $q_{\rm F}$ and $q$,
  and conjecture an exact formula for $q_{\rm F}=2$.
  However, for generic ($q_{\rm F},q$), the exact value of $d_{\rm F}$ remains unknown,
  and the exploration might request progresses in conformal field theory.

{\it Representations of the Potts model.} We begin with the introduction of the standard Potts, $q$-flow and RC models.
 Consider a finite graph $G \equiv (V,E)$,
where $V$ is the vertex set and $E$ the edge set.
 Let each vertex $i$ be occupied by a Potts spin
  $\sigma_i \! \in\! \{0, 1, \cdots, q-1\}$ with $q \! > \! 1$ an integer, the
$q$-state Potts model is defined by the probability distribution,
\begin{eqnarray}
  \!\!\!\!\!\! d \mu_{\rm spin} (\{\sigma \}) \!\!\! &=& \!\!\!{\cal Z}^{-1}_{\rm spin}  \!\prod_{(ij)} \exp \!\left[ J_{ij} (\delta_{\sigma_i,\sigma_j} \!-\!1) \right]\! d \mu_0  (\{\sigma \}) \nonumber
\end{eqnarray}
where $J_{ij} \! > \!  0$ is the ferromagnetic coupling for edge
$(ij) \! \in\! E$ in the graph $G$,
and $d \mu_0 (\{\sigma \})$ is the counting measure for the Potts spin
configurations. The partition sum $ {\cal Z}_{\rm spin}$ acts as a
normalization factor. Introducing the edge probability
  $p_{ij} \! \equiv \! 1 -  \exp(-J_{ij})$, the Potts distribution can be
  rewritten as,
\begin{eqnarray}
  \!\!\!\!\!\! d \mu_{\rm spin} (\{\sigma \}) \!\!\!  &=&  \!\!\!{\cal Z}^{-1}_{\rm spin}    \!\prod_{(ij)} \!\left[ p_{ij} \delta_{\sigma_i ,\sigma_j} \! + \! (1 \!- \! p_{ij})  \!\right]\! d \mu_0  (\{\sigma \}).
\label{eq:Potts}
\end{eqnarray}
Now, we can assign to each edge $(ij) \!  \in \!  E$ a random bond variable
$b_{ij} \! \in \! \{0,1\}$ and define the subgraph
$G_{\rm b } \!  \equiv \! (V, E_{\rm b} ) \subseteq G$, with $E_{\rm b}$ consisting of the
edges $(ij)$ with occupied bond $b_{ij} \! = \! 1$.  Let a cluster be a set of
vertices connected via occupied bonds, the constraint
$\delta_{\sigma_i ,\sigma_j}$ requires that all the Potts spins in the
same cluster take the same value, while the spin values in different
clusters are independent from each other.  After summing out the spin
degree of freedom, one obtains a FK bond configuration $\{ b \}$, in
which each cluster has a statistical weight of $q$.  The corresponding
RC model with parameter $q$ is then defined by the probability
distribution
\begin{equation}
  d \mu_{\rm FK} (\{b\}) \! = \! {\cal Z}^{-1}_{\rm FK} q^{k(G_{\rm b})}  \hspace{-3mm} \prod_{(ij) \in E_{\rm b}}  \hspace{-2mm} p_{ij}  \hspace{-2mm} \prod_{(ij) \not\in E_{\rm b} }  \hspace{-2mm} (1-p_{ij}) d \mu_0  (\{b \}) \hspace{0.2mm} , \label{eq:FK1}
\end{equation}
where $k(G_{\rm b})$ is the number of clusters in the graph
$G_{\rm b } $, including single-vertex clusters.

We can also add to each edge of $G$ a $q$-flow variable
$f_{ij} \! \in \! \{0,1,\cdots,q \! - \! 1\}$, and denote by
$G_{\rm f} \! \equiv \! (V, E_{\rm f}) \!  \subseteq \! G $ the subgraph of edges
$(ij)$ with nonzero flows $f_{ij} \! > \!0$.
Further, we introduce symbol $\partial G$ to represent the set of vertices
that do not satisfy the conservation condition given by the $q$-modular Kirchhoff conservation law  as
\begin{eqnarray}
\sum_{j{: (ij)\in E}} {\rm sgn}(i \rightarrow j) \; f_{ij} = 0 \mod q \; , \hspace{3mm} \mbox{for any } i \in V
  \label{eq:conservation}
\end{eqnarray}
where $\text{sgn}(i\rightarrow j)=-\text{sgn}(j\rightarrow i)\in\{\pm 1\}$ arises from the orientation of edge $(ij)$.
For any configuration $\{f\}$, the $q$-flow model is described by the probability
distribution,
\begin{align}
\!\! d \mu_{\rm qFlow} (\{f\}) &=  {\cal Z}^{-1}_{\rm qFlow} \; \delta_{\partial G= \varnothing} \notag \\
 &\times  \hspace{-2mm} \prod_{(ij) \in E_{\rm f}}  \hspace{-2mm} \frac{p_{ij}}{q} \hspace{-1.5mm} \prod_{(ij) \not\in E_{\rm f}} \hspace{-2mm}
 (1 \!- \!\tfrac{q-1}{q} p_{ij})  d \mu_0  (\{f \}) \; ,
\label{eq:qFlow}
\end{align}
where $\delta_{\partial G= \varnothing}$ means an empty set for $\partial G$, i.e., no vertex breaks the conservation law.
The orientation of each
  edge $(ij) \in E$ plays no physical role and can be randomly chosen,
  as reversing an edge $(ij)$ orientation can be counterbalanced by
  mapping the flow variable $f_{ij}$ to
  $q \! - \! f_{ij} \! \mod \! q$.

Using high-temperature expansion
\cite{Domb_1974,essam1986potts,wu1988potts,kasteleyn1969phase,fortuin1972random},
duality relations \cite{Wu_1982,caracciolo2004general} or
  low-temperature expansion for $2d$-planar graphs, it is known that
  ${\cal Z}_{\rm spin} = {\cal Z}_{\rm FK} = q^{|V|} {\cal Z}_{\rm
    qFlow}$ and, thus, apart from an unimportant constant $q^{|V|}$, the
  Potts~(\ref{eq:Potts}), RC~(\ref{eq:FK2}) and $q$-flow
  models~(\ref{eq:qFlow}) are equivalent to each other.

{\it Joint models.} In 1988, Edwards and Sokal defined a joint
model~\cite{edwards1988generalization}, having the $q$-state Potts
spin $\sigma_i$ at the vertices and occupation variable $b_{ij}$ on
the edges, with probability distribution
\begin{align}
d \mu_{\rm jSW} (\{\sigma\}, \{b\})&={\cal Z}^{-1}_{\rm jSW} \prod_{(ij)} [ p_{ij} \delta_{b_{ij},1} \delta_{\sigma_i, \sigma_j} \notag \\
&+(1\!-\!p_{ij}) \delta_{b_{ij},0} ] d \mu_0  (\{\sigma \}) d \mu_0  (\{b \}) \; .
\label{eq:jSW}
\end{align}
On this basis, the SW cluster algorithm can be easily understood as
passing back and forth between the spin and FK representations, via
the joint model~(\ref{eq:jSW}).  Given a spin configuration, a random
FK configuration is generated as follows: independently for each
edge $(ij)$, one sets $b_{ij} \! = \! 0$ for $\sigma_i \!  \neq \!  \sigma_j$, and
sets $b_{ij} \! = \! 1$ (resp. $0$) with probability $p_{ij}$
(resp. $(1-p_{ij})$), for $\sigma_i = \sigma_j$.  The reverse process
starts with a FK bond configuration. One picks equiprobably a $\sigma_i$
variable from the set $\{ 0, 1, \cdots, q \! -\! 1\}$ for each connected
cluster and assigns the $\sigma_i$ value to all the spins in this
cluster.

  \begin{figure*}
\includegraphics[width=0.9\textwidth]{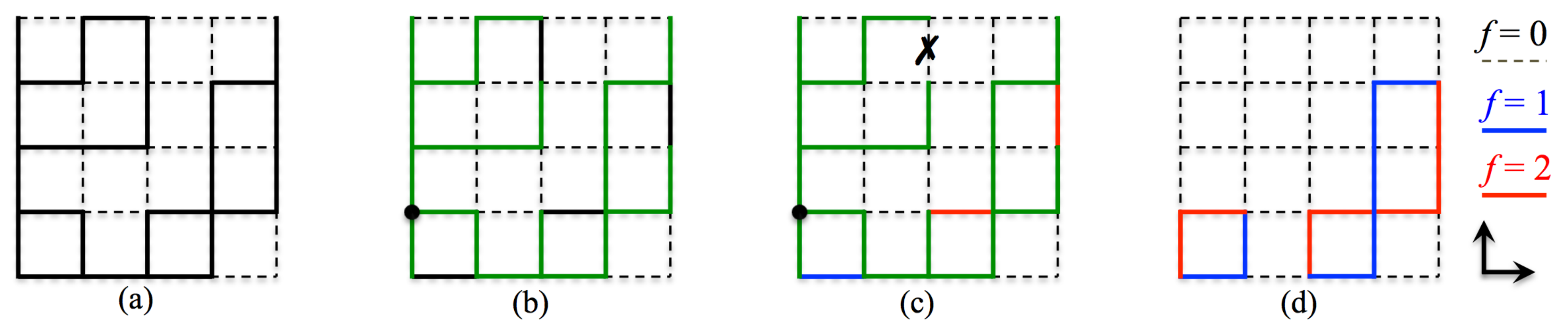}
\caption{Illustration of a cluster-to-loop update for $q=3$, with
    an up/right orientation, as shown in (d). From a FK bond
    configuration (a), a spanning tree is constructed from a root
    vertex, as marked by the green color (b). Each occupied edge missing
    from the tree defines an independent cycle and is assigned a
    random flow variable $f \in \{0,1,\cdots, q-1 \}$ (c).  Finally, the
    $q$-flow variables for all the other edges are obtained by
    backtracking vertices and applying the $q$-modular Kirchhoff
    conservation law for each vertex, yielding a $q$-flow
    configuration (d).}\label{fig:CL update}
\end{figure*}

We shall formulate a joint model between the FK bond and the
  $q$-flow configurations and the corresponding algorithm which passes
  back and forth. We first remark that, using the Euler formula
$ k(G_{\rm b}) = |V| - |E_{\rm b}| + c(G_{\rm b})$ where
$c(G_{\rm b})$ is the number of independent loops (cycles) in
$G_{\rm b}$, we can rewrite the RC model as
\begin{equation}
  d \mu_{\rm FK} (\{b\}) \! = \! {\cal Z}^{-1}_{\rm FK}  q^{|V|+c(G_{\rm b })}  \hspace{-3mm} \prod_{(ij) \in E_{\rm b}}  \hspace{-2mm} \frac{p_{ij}}{q}  \hspace{-2mm}   \prod_{(ij) \not\in E_{\rm b} }  \hspace{-2mm} \! (1\!-\!p_{ij}) d \mu_0  (\{b \}) \;,
\label{eq:FK2}
\end{equation}
which stresses the underlying cycle structure. Further, a simple
decomposition in the $q$-flow model leads to
\begin{align}
\!\! d \mu_{\rm qFlow} (\{f\}) &=  {\cal Z}^{-1}_{\rm qFlow} \; \delta_{\partial G= \varnothing} \notag \\
&\times  \hspace{-2mm} \prod_{(ij) \in E_{\rm f}}  \hspace{-2mm} \frac{p_{ij}}{q} \hspace{-1.5mm} \prod_{(ij) \not\in E_{\rm f}} \hspace{-2mm} (\frac{p_{ij}}{q}\!+\! 1\!-\!p_{ij})  d \mu_0  (\{f \}),
\label{eq:qFlow1}
\end{align}
as motivated by zero-valued flows corresponding modulo $q$
  to either $0$ or $q$ (resp. $1-p_{ij}$ and $p_{ij}/q$
  contributions). Analogously to \cite{edwards1988generalization}, we
define a joint model, having both the bond variable $b_{ij}$ and the
flow variable $f_{ij}$ on each edge, with the probability distribution
\begin{align}
d \mu_{\rm jLC} (\{ f\}, \{b\}) \!  &=  \!  {\cal Z}^{-1}_{\rm jLC} \;  \delta_{\partial G= \varnothing} \prod_{(ij)} \left[ \frac{p_{ij}}{q} \delta_{f_{ij} \neq 0} \delta_{b_{ij},1} \right. \notag \\
&+ \left. \frac{p_{ij}}{q} \delta_{f_{ij}= 0} \delta_{b_{ij},1}  +  (1-p_{ij}) \delta_{f_{ij}=0} \delta_{b_{ij},0} \right] \notag \\
&\times d \mu_0  (\{f \})  d \mu_0  (\{b \}) \; .
\label{eq:jLC}
\end{align}
We call this model the \emph{Loop-Cluster} (LC) joint model. As the
edge state $(f_{ij} \neq 0, b_{ij}= 0)$ is forbidden--i.e., has zero
probability, it yields $G_{\rm f} \subseteq G_{\rm b} \subseteq G$.
By explicitly performing the summation over either the $\{b \} $ or
the $\{f \} $ variables, it is easy to verify the following facts
about the LC joint model~(\ref{eq:jLC}):

(\romannumeral1) The marginal probability of the flow variables
$\{ f \}$ is precisely the $q$-flow model~(\ref{eq:qFlow1}), since,
  after summation over the bond states $b_{ij}=0,1$, an edge with the
  flow state $f_{ij} \neq 0$ has the statistical weight $\frac{p_{ij}}{q}$,
  and one with $f_{ij}=0$ a statistical weight of
  $(1-p_{ij})+\frac{p_{ij}}{q}$, as in (\ref{eq:qFlow1}).

  (\romannumeral2) The marginal probability of the bond variables
  $\{ b \}$ is precisely the RC model~(\ref{eq:FK2}).  The summation
  over the flow variables $\{ f \}$ involves the number of choices of
  assigning the flow variables under the constraints that
  $\partial G =\varnothing$ and the state
  $(f_{ij} \! \neq \! 0, b_{ij} \! =\! 0)$ is forbidden.  This number
  identifies with the number of possible flow configurations on the
  subgraph of occupied bonds, i.e. the flow configurations satisfying
  $\partial G_{\rm b} = \varnothing$. This number amounts to
  $q^{c(G_{\rm b})}$, by considering the decomposition of the
  Kirchhoff law~(\ref{eq:conservation}) into the loop flows on the
  graph $G_{\rm b}$.  Indeed, once the flow variable of an unshared
  edge of a loop is determined among the $q$ possible values, it must
  be propagated along the loop, defining the loop flow. The final flow
  for a given edge is the sum modulo $q$ of the loop flows it is
  contained in.  Thus, any bridge edge, i.e. not contained in any loop
  and whose removal would increase the number of clusters, is
  assigned a flow zero.

  (\romannumeral3) Given the flow variables $\{ f \}$, the bond
  variables $\{b\}$ are all independent and set by the conditional
  distribution $p(b_{ij}\!=\!1|f_{ij}\!>\!0)\! =\! 1$ for any edge
  $(i,j)$ with a non-zero flow and
  $p(b_{ij}\!=\!1|f_{ij}\!=\!0)\! =\!
  \frac{p_{ij}}{p_{ij}+q(1-p_{ij})} = t_{ij}$ otherwise.

  (\romannumeral4) Given the bond variables $\{ b \}$, the subset of flow
    variables $\{ f \}_{\rm b}$ on a cluster $G_{\rm b}$ is
    independent from the others and set by
    $p(\{f\}_b|G_{\rm b})\! =\! q^{-c(G_{\rm b})}\delta_{\partial
      G_{\rm b}=\varnothing}$
  and $p(f_{ij}\!=\!0|b_{ij}\!=\!0)=1$ for all edges $(ij)$ with unoccupied
  bonds.

  (\romannumeral5) The joint model (\ref{eq:jLC}) highlights the
  fundamental relationship between the FK and $q$-flow representations
  as both can be understood as the result of a high-temperature
  expansion over $\tfrac{p_{ij}}{1-p_{ij}}$ and $t_{ij}$,
  respectively, revealing either the connected-cluster or flow
  structure. Furthermore $t_{ij}$ identifies with the thermal
  transmissivity arising in the renormalization group
  \cite{wu1988potts,Tsallis_1981}.

{\it Loop-cluster algorithm.} We are now ready to formulate a LC Monte
Carlo method which simulates the joint model~(\ref{eq:jLC}).
% through a Gibbs/heat-bath sampling approach.  
To be specific, we alternatively
generate new bond variables, independent of the old ones, given the
flows following (\romannumeral3), and new flow variables, independent
of the old ones, given the bonds following (\romannumeral4).  The
marginal distribution $d \mu_{\rm FK}$ in~(\ref{eq:FK2})
($d \mu_{\rm qFlow} $ in~(\ref{eq:qFlow1})) from the joint
model~(\ref{eq:jLC}) is then simply obtained by erasing the flow
variables $\{f\}$ (bond variables $\{b\}$), as stated in
(\romannumeral1,\romannumeral2). This sampling procedure is a
generalization of the mapping method proposed in
\cite{grimmett2009random, Evertz_2002} for the Ising case.

(A) Given a $q$-flow configuration, generating a random FK bond
configuration is a straightforward local process given in
(\romannumeral3): for each non-zero flow $f_{ij} \! \neq \! 0$, one
sets $b_{ij} \! = \! 1$; for each edge with empty flow
$f_{ij} \! = \! 0$, one {\it independently} sets $b_{ij} \! = \! 1$
with probability $t_{ij}$, and $b_{ij} \! = \! 0$, otherwise.  The
number of operations in this step equals the number of edges of the
original graph, $|E|$.

(B) Given a FK bond configuration, generating a $q$-flow configuration
follows from (\romannumeral4) and depends on the subgraph-$G_{\rm b}$ topology:
For all the non-occupied edges $b_{ij}=0$, one sets $f_{ij}=0$;
  the edges in $E_{\rm b}$ are assigned  flow variables $\{ f\}$ as
  described in (\romannumeral2), once a set of independent loops have
  been defined.

  In more detail, we first construct a spanning tree for each
  connected cluster by a rooted procedure, either the breadth-first or
  the depth-first search.  Any occupied edge of the graph $G_{\rm b}$
  missing from the tree defines a loop by the symmetric difference of the
  tree paths from the pair of ending vertices of the missing edge to
  the root vertex.  Each of these occupied bonds is uniformly
  assigned a flow variable
  $f_{ij} \!  \in \! \{0, 1,\cdots, q \! - \! 1\}$.
  Then, we backtrack the tree and calculate the flow variables for all
  its edges by applying the $q$-modular Kirchhoff conservation law to
  each vertex.
  The number of operations is twice the number of edges in the original graph,
  $2|E|$.  Figure~\ref{fig:CL update} illustrates an example of
  ``constructing-tree" and ``backtracking" processes for $q=3$.
The number of operations is $3|E|$ for the LC scheme, slightly larger than $2|E|$ for the SW algorithm.

For $q=1$, the set of flow variables $\{0, \cdots, q-1\}$ reduces to $\{ 0 \}$
%and the zero-flow configuration $\{ f_{ij} \! = \! 0 \}$ becomes the only valid $q$-flow configuration.
and the LC algorithm becomes the conventional strategy for bond percolation.

The LC algorithm can be extended to sample from the RC model of real value $q \geq 1$,
 via the induced-subgraph decomposition~\cite{deng2007cluster}.
 Further, a single-cluster version can be formulated to sample from the $q$-flow model.
 See the supplementary material for details.

{\it Dynamical behavior.} We study numerically the dynamics
of the LC algorithm and compare it to the SW scheme for both
``energy-'' and ``susceptibility-like" quantities in the FK
representation at criticality over the complete graph and $d=2,3,4,5$
toroidals grids. By comparing the integrated autocorrelation times, we
obtain clear evidence that both the SW and LC schemes belong to the
same dynamical class (even displaying similar decorrelation performance
for $q=2$ in 2D), as well as the Wolff and the
single-cluster LC variant. Further details can be found in the
supplementary material.

{\it New family of fractal objects}.
The FK bond representation provides a platform to study rich geometric structures for any real $q \geq 0$. \
A variety of fractal dimensions are used to characterize the sizes of FK clusters,
the hulls, the external perimeters, the backbones and the shortest paths, etc.~\cite{stauffer2018introduction,kesten1987percolation}, and a set of exponents is
defined to account for correlation functions that two far-away regions are connected
by a number of mono- or polychromatic paths~\cite{smirnov2001critical,beffara2011monochromatic,aizenman1999path}.
% In the framework of logarithmic conformal field theory,
% the so-called $N$-cluster correlations ($N \geq 2$ is an integer) are recently constructed
% by using the $S_q$ symmetry of the Potts model~\cite{vasseur2012logarithmic,vasseur2014operator,couvreur2017non,tan2019observation}.
%In 2D, thanks to Coulomb-gas arguments~\cite{nienhuis1987two}, conformal field theory~\cite{cardy1987conformal} and stochastic Loewner evolution theory~\cite{lawler2001dimension},
In 2D, thanks to Coulomb-gas arguments, conformal field theory and stochastic Loewner evolution theory,
the exact values of most of these  exponents are available.
For instance,
one has the fractal dimension  $D_{\rm FK} = (g+2)(g+6)/8g$ for the FK clusters,
%$D_{\rm EP}=1+g/8$ for the external perimeters,
and the correlation exponent $X_2 = 1- 2/g$ for two polychromatic paths,
where the Coulomb-gas coupling $g \in [2,4]$ relates to $q$
as $q=2+2\cos(g\pi/2)$~\cite{Grossman_1987,Saleur_1987,Coniglio_1989}.
Nevertheless, exact values still remain unknown for a few exponents,
including the backbone dimension $D_{\rm bb}$.
For percolation ($q=1$), while the proximity of the numerical estimates for $D_{\rm bb}$ to the fraction $D_{\rm FK}-X_2 = 79/48 \approx  1.645\, 833 $  
has been noticed~\cite{Grassberger_1999,Huber}, this value seems ruled out by a high-precision study$D_{\rm bb}=1.643 \, 36(10)$~\cite{Xu_2014}.

 As in the FK representation, clusters can be defined as sets of vertices connected
  via edges of non-zero flows in a $q$-flow configuration, which have so far received little attention.
  From the LC joint model~(\ref{eq:jLC}), it is seen that a FK cluster may contain more than
  one $q$-flow cluster while the reverse cannot occur.
  Actually, since any bridge edge has a zero flow, the
  $q$-flow clusters must live on top of the backbones of the FK clusters--sets of vertices connected via non-bridge edges.
  In practice, since any loop has a flow zero with probability
   $1/q$, $q$-flow clusters are generally smaller than the backbone
   clusters and, therefore, one has
   $D_{\rm qF} \! \leq \! D_{\rm bb} \! \leq \! D_{\rm FK}$.

 \begin{figure}
\includegraphics[width=1.0\columnwidth]{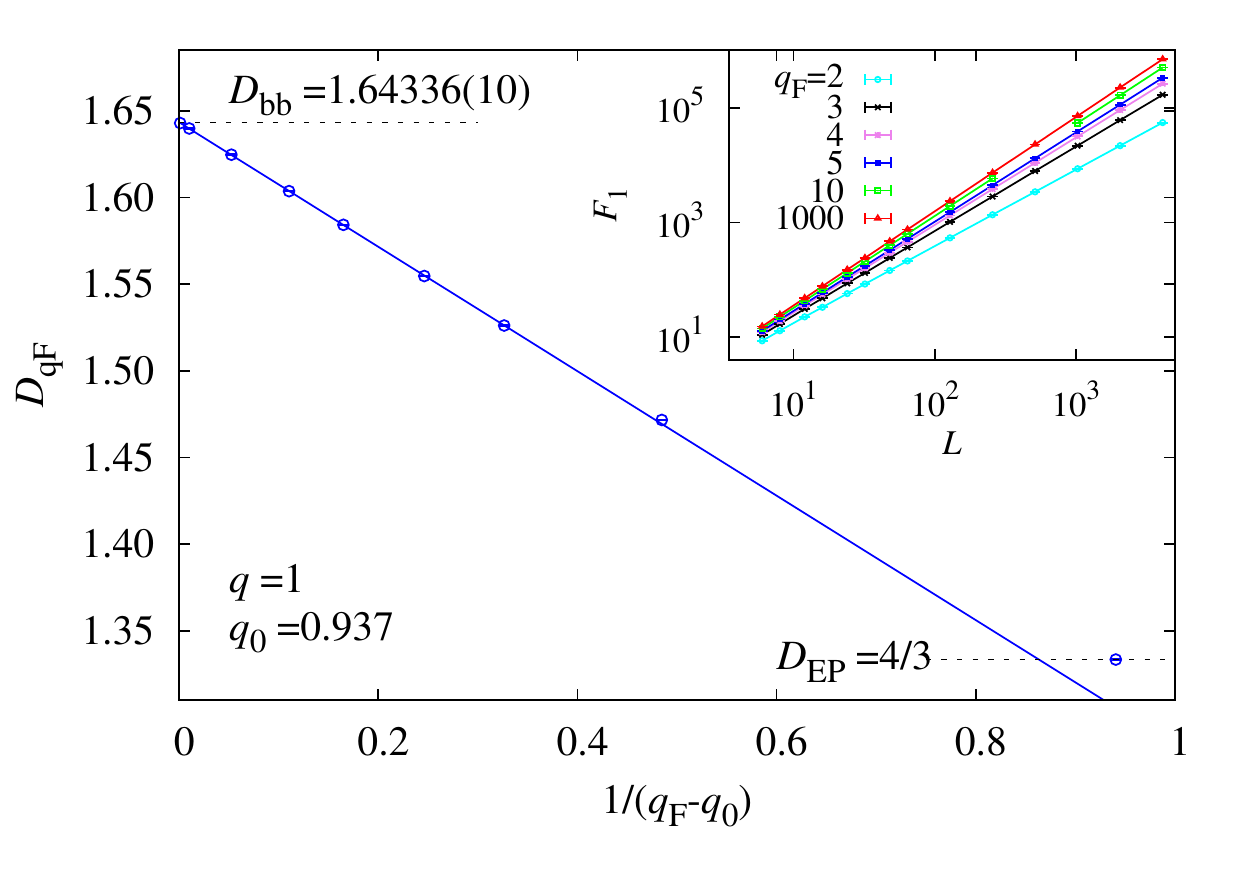}
\caption{Convergence of the fractal dimension $D_{\rm qF}$ of
  $q_{\rm F}$-flow clusters from $D_{\rm EP} = 4/3$ to the backbone
  dimension $D_{\rm bb}=1.64336$ with $1/(q_{\rm F}-q_0), q_0=0.937$ for the 2D percolation.
  Inset: Scaling of the  size $F_1$ of the largest $q_{\rm F}$-flow clusters for increasing value of $q_{\rm F}$.}
  \label{fig:fractal}
\end{figure}

  Further, given a $q$-state FK bond configuration, we can introduce an integer parameter
  $q_{\rm F}\geq 2$ such that, in Step B of the LC scheme for assigning flow variables,
  each loop has a flow zero with probability $1/q_{\rm F}$
  and the $q_{\rm F}$-modular conservation law applies to  each vertex.
  This leads to a hierarchy of $q_{\rm F}$-flow clusters,
  reducing to the $q$-flow clusters for $q_{\rm F} = q$.
  Note that Step A  can no longer be applied if $q_{\rm F} \neq q$, and
  the FK configuration has to be updated by other means like the cluster or Sweeny algorithms~\cite{sweeny1983monte,swendsen1987nonuniversal,wolff1989collective}.

  We carry out extensive simulation for ($q=1,2,3, 2+\sqrt{3}, q_{\rm F}=2$) and ($q=1, q_{\rm F}=2,3,4,5,7,10,20,100,1000$) on the 2D-toroidal grid
  with linear size $L \in [6,4096]$.
  From finite-size scaling analysis, we determine the fractal dimension $D_{\rm qF}$
  for the $q_{\rm F}$-flow clusters.
  For $q_{\rm F}=2$, the results are $D_{\rm qF} = 1.333\, 3(2) \approx 4/3$, $1.375 \, 4 (12) \approx 11/8$,
  $1.417 (2) \approx 17/12$ and $1.464 (6) \approx 35/24$ for $q=1,2,3,2+\sqrt{3}$, respectively.
  These are well consistent with the external-perimeter fractal dimension $D_{\rm EP}=1+g/8$~\cite{Saleur_1987},
  and, thus,  we conjecture $D_{\rm qF} (q_{\rm F}=2) = D_{\rm EP}$.

  For percolation ($q \! = \! 1$), we obtain $D_{\rm qF} \! = \! 1.471\, 6(2)$, $1.526 \, 1(2)$,
  $1.554 \, 7 (2)$, $1.584\,2(2)$, $1.603\,6(2)$, $1.624\,7(2)$, $1.639 \, 8(2)$ and $1.642 \, 9(2)$
  for $q_{\rm F}=3$,4,5,7,10,20,100,1000, respectively.
  As $q_{\rm F}$ increases, $D_{\rm qF}$ converges to the backbone dimension
  asymptotically as $1/(q_{\rm F}-q_0)$.
  A least-squares fit with $q_{\rm F} \geq 4$ yields $q_0=0.94 (4)$ and $D_{\rm qF} (q_{\rm F} \! \rightarrow \! \infty)  \! = \! 1.643 \, 4(2)$,
  which agrees well with $D_{\rm bb}=1.643 \, 36(10)$~\cite{Xu_2014}.

{\it Conclusion.} We introduce the LC joint model of the FK bond and
$q$-flow representations of the Potts model, unifying its three standard representations.
A straightforward application is the design of LC algorithms.
While in the same dynamical class as the SW and Wolff methods,
the LC algorithms lift the limitation of performing both simulations and measurements in a given representation.
More importantly, the LC coupling sheds  much new light on the geometric properties of
FK and $q$-flow clusters.
It is proved that the $q$-flow clusters have a fractal dimension not larger than the
 backbone one of the FK clusters.
 Further, a hierarchy of $q_{\rm{F}}$-flow clusters is constructed with integer $q_{\rm F} \geq 2$,
 enriching the characterization of fractal structures of the FK clusters.
 In two dimensions, from our high-precision results
 we conjecture $D_{\rm qF} (q_{\rm F} \! = \! 2) \! = \! D_{\rm EP} \! = \! 1+g/8$;
 otherwise, the exact values of $D_{\rm qF}$ are not available for generic $(q,q_{\rm F})$.
 Future works shall focus on an extensive study in the $(q,q_{\rm F})$ diagram
 and seek for the exact formula of $D_{\rm qF}$ in two dimensions.

 \begin{acknowledgments}
 We dedicate this work to Fred (Fa-Yueh) Wu who passed away on January 21, 2020. 
 His seminal review article on the Potts model~\cite{Wu_1982}
has benefitted generations of statistical physicists, and he was one of the early researchers who paid attention 
 to the flow representation of the Potts model~\cite{wu1988potts}.
Wu was a member of the doctoral dissertation
committee of one of us (Y.D.) in 2004, and subsequently gave him a lot
of encouragement throughout his academic career.
   This work was supported by the Ministry of Science and Technology
   of China for Grant No.~2016YFA0301604 and the National Natural
   Science Foundation of China for Grant No.~11625522. M. Michel is
   grateful for the support of the PHC program Xu Guangqi (Grant
   No. 41291UF). We thank Shanglun Feng and Ziming Cheng for their
   early involvements in the work.
 \end{acknowledgments}
\vfill
\bibliographystyle{apsrev4-1}
\bibliography{reference}

\begin{thebibliography}{55}%
\makeatletter
\providecommand \@ifxundefined [1]{%
 \@ifx{#1\undefined}
}%
\providecommand \@ifnum [1]{%
 \ifnum #1\expandafter \@firstoftwo
 \else \expandafter \@secondoftwo
 \fi
}%
\providecommand \@ifx [1]{%
 \ifx #1\expandafter \@firstoftwo
 \else \expandafter \@secondoftwo
 \fi
}%
\providecommand \natexlab [1]{#1}%
\providecommand \enquote  [1]{``#1''}%
\providecommand \bibnamefont  [1]{#1}%
\providecommand \bibfnamefont [1]{#1}%
\providecommand \citenamefont [1]{#1}%
\providecommand \href@noop [0]{\@secondoftwo}%
\providecommand \href [0]{\begingroup \@sanitize@url \@href}%
\providecommand \@href[1]{\@@startlink{#1}\@@href}%
\providecommand \@@href[1]{\endgroup#1\@@endlink}%
\providecommand \@sanitize@url [0]{\catcode `\\12\catcode `\$12\catcode
  `\&12\catcode `\#12\catcode `\^12\catcode `\_12\catcode `\%12\relax}%
\providecommand \@@startlink[1]{}%
\providecommand \@@endlink[0]{}%
\providecommand \url  [0]{\begingroup\@sanitize@url \@url }%
\providecommand \@url [1]{\endgroup\@href {#1}{\urlprefix }}%
\providecommand \urlprefix  [0]{URL }%
\providecommand \Eprint [0]{\href }%
\providecommand \doibase [0]{http://dx.doi.org/}%
\providecommand \selectlanguage [0]{\@gobble}%
\providecommand \bibinfo  [0]{\@secondoftwo}%
\providecommand \bibfield  [0]{\@secondoftwo}%
\providecommand \translation [1]{[#1]}%
\providecommand \BibitemOpen [0]{}%
\providecommand \bibitemStop [0]{}%
\providecommand \bibitemNoStop [0]{.\EOS\space}%
\providecommand \EOS [0]{\spacefactor3000\relax}%
\providecommand \BibitemShut  [1]{\csname bibitem#1\endcsname}%
\let\auto@bib@innerbib\@empty
%</preamble>
\bibitem [{\citenamefont {Wu}(1982)}]{Wu_1982}%
  \BibitemOpen
  \bibfield  {author} {\bibinfo {author} {\bibfnamefont {F.~Y.}\ \bibnamefont
  {Wu}},\ }\href {\doibase 10.1103/RevModPhys.54.235} {\bibfield  {journal}
  {\bibinfo  {journal} {Rev. Mod. Phys.}\ }\textbf {\bibinfo {volume} {54}},\
  \bibinfo {pages} {235} (\bibinfo {year} {1982})}\BibitemShut {NoStop}%
\bibitem [{\citenamefont {Baxter}(1989)}]{baxter2016exactly}%
  \BibitemOpen
  \bibfield  {author} {\bibinfo {author} {\bibfnamefont {R.~J.}\ \bibnamefont
  {Baxter}},\ }\href
  {https://www.elsevier.com/books/exactly-solved-models-in-statistical-mechanics/baxter/978-0-12-083182-1}
  {\emph {\bibinfo {title} {Exactly solved models in statistical mechanics}}}\
  (\bibinfo  {publisher} {Elsevier},\ \bibinfo {year} {1989})\BibitemShut
  {NoStop}%
\bibitem [{\citenamefont {Nienhuis}(1984)}]{nienhuis1984critical}%
  \BibitemOpen
  \bibfield  {author} {\bibinfo {author} {\bibfnamefont {B.}~\bibnamefont
  {Nienhuis}},\ }\href {\doibase 10.1007/BF01009437} {\bibfield  {journal}
  {\bibinfo  {journal} {J. Stat. Phys.}\ }\textbf {\bibinfo {volume} {34}},\
  \bibinfo {pages} {731} (\bibinfo {year} {1984})}\BibitemShut {NoStop}%
\bibitem [{\citenamefont {Essam}\ and\ \citenamefont
  {Tsallis}(1986)}]{essam1986potts}%
  \BibitemOpen
  \bibfield  {author} {\bibinfo {author} {\bibfnamefont {J.}~\bibnamefont
  {Essam}}\ and\ \bibinfo {author} {\bibfnamefont {C.}~\bibnamefont
  {Tsallis}},\ }\href {\doibase 10.1088/0305-4470/19/3/022} {\bibfield
  {journal} {\bibinfo  {journal} {J. Phys. A}\ }\textbf {\bibinfo {volume}
  {19}},\ \bibinfo {pages} {409} (\bibinfo {year} {1986})}\BibitemShut
  {NoStop}%
\bibitem [{\citenamefont {Wu}(1988)}]{wu1988potts}%
  \BibitemOpen
  \bibfield  {author} {\bibinfo {author} {\bibfnamefont {F.~Y.}\ \bibnamefont
  {Wu}},\ }\href {\doibase 10.1007/BF01016406} {\bibfield  {journal} {\bibinfo
  {journal} {J. Stat. Phys.}\ }\textbf {\bibinfo {volume} {52}},\ \bibinfo
  {pages} {99} (\bibinfo {year} {1988})}\BibitemShut {NoStop}%
\bibitem [{\citenamefont {Kasteleyn}\ and\ \citenamefont
  {Fortuin}(1969)}]{kasteleyn1969phase}%
  \BibitemOpen
  \bibfield  {author} {\bibinfo {author} {\bibfnamefont {P.}~\bibnamefont
  {Kasteleyn}}\ and\ \bibinfo {author} {\bibfnamefont {C.}~\bibnamefont
  {Fortuin}},\ }\href@noop {} {\bibfield  {journal} {\bibinfo  {journal} {J.
  Phys. Soc. Jpn. Suppl.}\ }\textbf {\bibinfo {volume} {26}},\ \bibinfo {pages}
  {11} (\bibinfo {year} {1969})}\BibitemShut {NoStop}%
\bibitem [{\citenamefont {Fortuin}\ and\ \citenamefont
  {Kasteleyn}(1972)}]{fortuin1972random}%
  \BibitemOpen
  \bibfield  {author} {\bibinfo {author} {\bibfnamefont {C.~M.}\ \bibnamefont
  {Fortuin}}\ and\ \bibinfo {author} {\bibfnamefont {P.~W.}\ \bibnamefont
  {Kasteleyn}},\ }\href {\doibase 10.1016/0031-8914(72)90045-6} {\bibfield
  {journal} {\bibinfo  {journal} {Physica}\ }\textbf {\bibinfo {volume} {57}},\
  \bibinfo {pages} {536} (\bibinfo {year} {1972})}\BibitemShut {NoStop}%
\bibitem [{\citenamefont {Chayes}\ and\ \citenamefont
  {Machta}(1998)}]{chayes1998graphical}%
  \BibitemOpen
  \bibfield  {author} {\bibinfo {author} {\bibfnamefont {L.}~\bibnamefont
  {Chayes}}\ and\ \bibinfo {author} {\bibfnamefont {J.}~\bibnamefont
  {Machta}},\ }\href {\doibase 10.1016/S0378-4371(97)00637-7} {\bibfield
  {journal} {\bibinfo  {journal} {Physica A}\ }\textbf {\bibinfo {volume}
  {254}},\ \bibinfo {pages} {477} (\bibinfo {year} {1998})}\BibitemShut
  {NoStop}%
\bibitem [{\citenamefont {Deng}\ \emph
  {et~al.}(2007{\natexlab{a}})\citenamefont {Deng}, \citenamefont {Garoni},
  \citenamefont {Guo}, \citenamefont {Bl{\"o}te},\ and\ \citenamefont
  {Sokal}}]{deng2007cluster}%
  \BibitemOpen
  \bibfield  {author} {\bibinfo {author} {\bibfnamefont {Y.}~\bibnamefont
  {Deng}}, \bibinfo {author} {\bibfnamefont {T.~M.}\ \bibnamefont {Garoni}},
  \bibinfo {author} {\bibfnamefont {W.}~\bibnamefont {Guo}}, \bibinfo {author}
  {\bibfnamefont {H.~W.}\ \bibnamefont {Bl{\"o}te}}, \ and\ \bibinfo {author}
  {\bibfnamefont {A.~D.}\ \bibnamefont {Sokal}},\ }\href {\doibase
  10.1103/PhysRevLett.98.120601} {\bibfield  {journal} {\bibinfo  {journal}
  {Phys. Rev. Lett.}\ }\textbf {\bibinfo {volume} {98}},\ \bibinfo {pages}
  {120601} (\bibinfo {year} {2007}{\natexlab{a}})}\BibitemShut {NoStop}%
\bibitem [{\citenamefont {Deng}\ \emph
  {et~al.}(2007{\natexlab{b}})\citenamefont {Deng}, \citenamefont {Garoni},
  \citenamefont {Machta}, \citenamefont {Ossola}, \citenamefont {Polin},\ and\
  \citenamefont {Sokal}}]{deng2007critical}%
  \BibitemOpen
  \bibfield  {author} {\bibinfo {author} {\bibfnamefont {Y.}~\bibnamefont
  {Deng}}, \bibinfo {author} {\bibfnamefont {T.~M.}\ \bibnamefont {Garoni}},
  \bibinfo {author} {\bibfnamefont {J.}~\bibnamefont {Machta}}, \bibinfo
  {author} {\bibfnamefont {G.}~\bibnamefont {Ossola}}, \bibinfo {author}
  {\bibfnamefont {M.}~\bibnamefont {Polin}}, \ and\ \bibinfo {author}
  {\bibfnamefont {A.~D.}\ \bibnamefont {Sokal}},\ }\href {\doibase
  10.1103/PhysRevLett.99.055701} {\bibfield  {journal} {\bibinfo  {journal}
  {Phys. Rev. Lett.}\ }\textbf {\bibinfo {volume} {99}},\ \bibinfo {pages}
  {055701} (\bibinfo {year} {2007}{\natexlab{b}})}\BibitemShut {NoStop}%
\bibitem [{\citenamefont {Di~Francesco}\ \emph {et~al.}(1997)\citenamefont
  {Di~Francesco}, \citenamefont {Mathieu},\ and\ \citenamefont
  {Senechal}}]{di1997conformal}%
  \BibitemOpen
  \bibfield  {author} {\bibinfo {author} {\bibfnamefont {P.}~\bibnamefont
  {Di~Francesco}}, \bibinfo {author} {\bibfnamefont {P.}~\bibnamefont
  {Mathieu}}, \ and\ \bibinfo {author} {\bibfnamefont {D.}~\bibnamefont
  {Senechal}},\ }\href {\doibase 10.1007/978-1-4612-2256-9} {\emph {\bibinfo
  {title} {Conformal field theory}}}\ (\bibinfo  {publisher} {Springer},\
  \bibinfo {address} {New York},\ \bibinfo {year} {1997})\BibitemShut {NoStop}%
\bibitem [{\citenamefont {Schramm}(2000)}]{schramm2000schramm}%
  \BibitemOpen
  \bibfield  {author} {\bibinfo {author} {\bibfnamefont {O.}~\bibnamefont
  {Schramm}},\ }\href {\doibase 10.1007/BF02803524} {\bibfield  {journal}
  {\bibinfo  {journal} {Isr. J. Math.}\ }\textbf {\bibinfo {volume} {118}},\
  \bibinfo {pages} {221} (\bibinfo {year} {2000})}\BibitemShut {NoStop}%
\bibitem [{\citenamefont {Rohde}\ and\ \citenamefont
  {Schramm}(2005)}]{rohde2005s}%
  \BibitemOpen
  \bibfield  {author} {\bibinfo {author} {\bibfnamefont {S.}~\bibnamefont
  {Rohde}}\ and\ \bibinfo {author} {\bibfnamefont {O.}~\bibnamefont
  {Schramm}},\ }\href {\doibase 10.4007/annals.2005.161.883} {\bibfield
  {journal} {\bibinfo  {journal} {Ann. Math.}\ }\textbf {\bibinfo {volume}
  {161}},\ \bibinfo {pages} {883} (\bibinfo {year} {2005})}\BibitemShut
  {NoStop}%
\bibitem [{\citenamefont {Lawler}(2005)}]{lawler2005conformally}%
  \BibitemOpen
  \bibfield  {author} {\bibinfo {author} {\bibfnamefont {G.~F.}\ \bibnamefont
  {Lawler}},\ }\href {https://bookstore.ams.org/surv-114-s} {\emph {\bibinfo
  {title} {Conformally invariant processes in the plane}}},\ \bibinfo {number}
  {114}\ (\bibinfo  {publisher} {American Mathematical Soc.},\ \bibinfo {year}
  {2005})\BibitemShut {NoStop}%
\bibitem [{\citenamefont {Kager}\ and\ \citenamefont
  {Nienhuis}(2004)}]{kager2004guide}%
  \BibitemOpen
  \bibfield  {author} {\bibinfo {author} {\bibfnamefont {W.}~\bibnamefont
  {Kager}}\ and\ \bibinfo {author} {\bibfnamefont {B.}~\bibnamefont
  {Nienhuis}},\ }\href {\doibase 10.1023/B:JOSS.0000028058.87266.be} {\bibfield
   {journal} {\bibinfo  {journal} {J. Stat. Phys.}\ }\textbf {\bibinfo {volume}
  {115}},\ \bibinfo {pages} {1149} (\bibinfo {year} {2004})}\BibitemShut
  {NoStop}%
\bibitem [{\citenamefont {Cardy}(2005)}]{cardy2005sle}%
  \BibitemOpen
  \bibfield  {author} {\bibinfo {author} {\bibfnamefont {J.}~\bibnamefont
  {Cardy}},\ }\href {\doibase 10.1016/j.aop.2005.04.001} {\bibfield  {journal}
  {\bibinfo  {journal} {Ann. Phys.}\ }\textbf {\bibinfo {volume} {318}},\
  \bibinfo {pages} {81} (\bibinfo {year} {2005})}\BibitemShut {NoStop}%
\bibitem [{\citenamefont {Metropolis}\ \emph {et~al.}(1953)\citenamefont
  {Metropolis}, \citenamefont {Rosenbluth}, \citenamefont {Rosenbluth},
  \citenamefont {Teller},\ and\ \citenamefont
  {Teller}}]{metropolis1953equation}%
  \BibitemOpen
  \bibfield  {author} {\bibinfo {author} {\bibfnamefont {N.}~\bibnamefont
  {Metropolis}}, \bibinfo {author} {\bibfnamefont {A.~W.}\ \bibnamefont
  {Rosenbluth}}, \bibinfo {author} {\bibfnamefont {M.~N.}\ \bibnamefont
  {Rosenbluth}}, \bibinfo {author} {\bibfnamefont {A.~H.}\ \bibnamefont
  {Teller}}, \ and\ \bibinfo {author} {\bibfnamefont {E.}~\bibnamefont
  {Teller}},\ }\href {\doibase 10.1063/1.1699114} {\bibfield  {journal}
  {\bibinfo  {journal} {J. Chem. Phys.}\ }\textbf {\bibinfo {volume} {21}},\
  \bibinfo {pages} {1087} (\bibinfo {year} {1953})}\BibitemShut {NoStop}%
\bibitem [{\citenamefont {Hohenberg}\ and\ \citenamefont
  {Halperin}(1977)}]{hohenberg1977theory}%
  \BibitemOpen
  \bibfield  {author} {\bibinfo {author} {\bibfnamefont {P.~C.}\ \bibnamefont
  {Hohenberg}}\ and\ \bibinfo {author} {\bibfnamefont {B.~I.}\ \bibnamefont
  {Halperin}},\ }\href {\doibase 10.1103/RevModPhys.49.435} {\bibfield
  {journal} {\bibinfo  {journal} {Rev. Mod. Phys.}\ }\textbf {\bibinfo {volume}
  {49}},\ \bibinfo {pages} {435} (\bibinfo {year} {1977})}\BibitemShut
  {NoStop}%
\bibitem [{\citenamefont {Sokal}(1997)}]{sokal1997functional}%
  \BibitemOpen
  \bibfield  {author} {\bibinfo {author} {\bibfnamefont {A.}~\bibnamefont
  {Sokal}},\ }\enquote {\bibinfo {title} {Monte carlo methods in statistical
  mechanics: Foundations and new algorithms},}\ in\ \href {\doibase
  10.1007/978-1-4899-0319-8_6} {\emph {\bibinfo {booktitle} {Functional
  Integration: Basics and Applications}}},\ \bibinfo {editor} {edited by\
  \bibinfo {editor} {\bibfnamefont {C.}~\bibnamefont {DeWitt-Morette}},
  \bibinfo {editor} {\bibfnamefont {P.}~\bibnamefont {Cartier}}, \ and\
  \bibinfo {editor} {\bibfnamefont {A.}~\bibnamefont {Folacci}}}\ (\bibinfo
  {publisher} {Springer US},\ \bibinfo {address} {Boston, MA},\ \bibinfo {year}
  {1997})\ pp.\ \bibinfo {pages} {131--192}\BibitemShut {NoStop}%
\bibitem [{\citenamefont {Sweeny}(1983)}]{sweeny1983monte}%
  \BibitemOpen
  \bibfield  {author} {\bibinfo {author} {\bibfnamefont {M.}~\bibnamefont
  {Sweeny}},\ }\href {\doibase 10.1103/PhysRevB.27.4445} {\bibfield  {journal}
  {\bibinfo  {journal} {Phys. Rev. B}\ }\textbf {\bibinfo {volume} {27}},\
  \bibinfo {pages} {4445} (\bibinfo {year} {1983})}\BibitemShut {NoStop}%
\bibitem [{\citenamefont {Edwards}\ and\ \citenamefont
  {Sokal}(1988)}]{edwards1988generalization}%
  \BibitemOpen
  \bibfield  {author} {\bibinfo {author} {\bibfnamefont {R.~G.}\ \bibnamefont
  {Edwards}}\ and\ \bibinfo {author} {\bibfnamefont {A.~D.}\ \bibnamefont
  {Sokal}},\ }\href {\doibase 10.1103/PhysRevD.38.2009} {\bibfield  {journal}
  {\bibinfo  {journal} {Phys. Rev. D}\ }\textbf {\bibinfo {volume} {38}},\
  \bibinfo {pages} {2009} (\bibinfo {year} {1988})}\BibitemShut {NoStop}%
\bibitem [{\citenamefont {Swendsen}\ and\ \citenamefont
  {Wang}(1987)}]{swendsen1987nonuniversal}%
  \BibitemOpen
  \bibfield  {author} {\bibinfo {author} {\bibfnamefont {R.~H.}\ \bibnamefont
  {Swendsen}}\ and\ \bibinfo {author} {\bibfnamefont {J.-S.}\ \bibnamefont
  {Wang}},\ }\href {\doibase 10.1103/PhysRevLett.58.86} {\bibfield  {journal}
  {\bibinfo  {journal} {Phys. Rev. Lett.}\ }\textbf {\bibinfo {volume} {58}},\
  \bibinfo {pages} {86} (\bibinfo {year} {1987})}\BibitemShut {NoStop}%
\bibitem [{\citenamefont {Wolff}(1989)}]{wolff1989collective}%
  \BibitemOpen
  \bibfield  {author} {\bibinfo {author} {\bibfnamefont {U.}~\bibnamefont
  {Wolff}},\ }\href {\doibase 10.1103/PhysRevLett.62.361} {\bibfield  {journal}
  {\bibinfo  {journal} {Phys. Rev. Lett.}\ }\textbf {\bibinfo {volume} {62}},\
  \bibinfo {pages} {361} (\bibinfo {year} {1989})}\BibitemShut {NoStop}%
\bibitem [{\citenamefont {Prokof'ev}\ and\ \citenamefont
  {Svistunov}(2001)}]{prokof2001worm}%
  \BibitemOpen
  \bibfield  {author} {\bibinfo {author} {\bibfnamefont {N.}~\bibnamefont
  {Prokof'ev}}\ and\ \bibinfo {author} {\bibfnamefont {B.}~\bibnamefont
  {Svistunov}},\ }\href {\doibase 10.1103/PhysRevLett.87.160601} {\bibfield
  {journal} {\bibinfo  {journal} {Phys. Rev. Lett.}\ }\textbf {\bibinfo
  {volume} {87}},\ \bibinfo {pages} {160601} (\bibinfo {year}
  {2001})}\BibitemShut {NoStop}%
\bibitem [{\citenamefont {Prokof'ev}\ \emph {et~al.}(1998)\citenamefont
  {Prokof'ev}, \citenamefont {Svistunov},\ and\ \citenamefont
  {Tupitsyn}}]{prokof1998worm}%
  \BibitemOpen
  \bibfield  {author} {\bibinfo {author} {\bibfnamefont {N.}~\bibnamefont
  {Prokof'ev}}, \bibinfo {author} {\bibfnamefont {B.}~\bibnamefont
  {Svistunov}}, \ and\ \bibinfo {author} {\bibfnamefont {I.}~\bibnamefont
  {Tupitsyn}},\ }\href {\doibase 10.1016/S0375-9601(97)00957-2} {\bibfield
  {journal} {\bibinfo  {journal} {Phys. Lett. A}\ }\textbf {\bibinfo {volume}
  {238}},\ \bibinfo {pages} {253} (\bibinfo {year} {1998})}\BibitemShut
  {NoStop}%
\bibitem [{\citenamefont {Mercado}\ \emph {et~al.}(2012)\citenamefont
  {Mercado}, \citenamefont {Evertz},\ and\ \citenamefont
  {Gattringer}}]{mercado2012worm}%
  \BibitemOpen
  \bibfield  {author} {\bibinfo {author} {\bibfnamefont {Y.~D.}\ \bibnamefont
  {Mercado}}, \bibinfo {author} {\bibfnamefont {H.~G.}\ \bibnamefont {Evertz}},
  \ and\ \bibinfo {author} {\bibfnamefont {C.}~\bibnamefont {Gattringer}},\
  }\href {\doibase 10.1016/j.cpc.2012.04.014} {\bibfield  {journal} {\bibinfo
  {journal} {Comput. Phys. Commun.}\ }\textbf {\bibinfo {volume} {183}},\
  \bibinfo {pages} {1920} (\bibinfo {year} {2012})}\BibitemShut {NoStop}%
\bibitem [{\citenamefont {El{\c{c}}i}\ \emph {et~al.}(2018)\citenamefont
  {El{\c{c}}i}, \citenamefont {Grimm}, \citenamefont {Ding}, \citenamefont
  {Nasrawi}, \citenamefont {Garoni},\ and\ \citenamefont
  {Deng}}]{elcci2018lifted}%
  \BibitemOpen
  \bibfield  {author} {\bibinfo {author} {\bibfnamefont {E.~M.}\ \bibnamefont
  {El{\c{c}}i}}, \bibinfo {author} {\bibfnamefont {J.}~\bibnamefont {Grimm}},
  \bibinfo {author} {\bibfnamefont {L.}~\bibnamefont {Ding}}, \bibinfo {author}
  {\bibfnamefont {A.}~\bibnamefont {Nasrawi}}, \bibinfo {author} {\bibfnamefont
  {T.~M.}\ \bibnamefont {Garoni}}, \ and\ \bibinfo {author} {\bibfnamefont
  {Y.}~\bibnamefont {Deng}},\ }\href {\doibase 10.1103/PhysRevE.97.042126}
  {\bibfield  {journal} {\bibinfo  {journal} {Phys. Rev. E}\ }\textbf {\bibinfo
  {volume} {97}},\ \bibinfo {pages} {042126} (\bibinfo {year}
  {2018})}\BibitemShut {NoStop}%
\bibitem [{\citenamefont {Deng}\ \emph
  {et~al.}(2007{\natexlab{c}})\citenamefont {Deng}, \citenamefont {Garoni},\
  and\ \citenamefont {Sokal}}]{deng2007dynamic}%
  \BibitemOpen
  \bibfield  {author} {\bibinfo {author} {\bibfnamefont {Y.}~\bibnamefont
  {Deng}}, \bibinfo {author} {\bibfnamefont {T.~M.}\ \bibnamefont {Garoni}}, \
  and\ \bibinfo {author} {\bibfnamefont {A.~D.}\ \bibnamefont {Sokal}},\ }\href
  {\doibase 10.1103/PhysRevLett.99.110601} {\bibfield  {journal} {\bibinfo
  {journal} {Phys. Rev. Lett.}\ }\textbf {\bibinfo {volume} {99}},\ \bibinfo
  {pages} {110601} (\bibinfo {year} {2007}{\natexlab{c}})}\BibitemShut
  {NoStop}%
\bibitem [{\citenamefont {Wolff}(2009)}]{wolff2009simulating}%
  \BibitemOpen
  \bibfield  {author} {\bibinfo {author} {\bibfnamefont {U.}~\bibnamefont
  {Wolff}},\ }\href {\doibase 10.1016/j.nuclphysb.2008.09.033} {\bibfield
  {journal} {\bibinfo  {journal} {Nucl. Phys. B}\ }\textbf {\bibinfo {volume}
  {810}},\ \bibinfo {pages} {491} (\bibinfo {year} {2009})}\BibitemShut
  {NoStop}%
\bibitem [{\citenamefont {Grimmett}\ and\ \citenamefont
  {Janson}(2009)}]{grimmett2009random}%
  \BibitemOpen
  \bibfield  {author} {\bibinfo {author} {\bibfnamefont {G.}~\bibnamefont
  {Grimmett}}\ and\ \bibinfo {author} {\bibfnamefont {S.}~\bibnamefont
  {Janson}},\ }\href
  {https://www.combinatorics.org/ojs/index.php/eljc/article/view/v16i1r46}
  {\bibfield  {journal} {\bibinfo  {journal} {Electron. J. Comb.}\ }\textbf
  {\bibinfo {volume} {16}},\ \bibinfo {pages} {R46} (\bibinfo {year}
  {2009})}\BibitemShut {NoStop}%
\bibitem [{\citenamefont {Evertz}\ \emph {et~al.}(2002)\citenamefont {Evertz},
  \citenamefont {Erkinger},\ and\ \citenamefont {{Von der
  Linden}}}]{Evertz_2002}%
  \BibitemOpen
  \bibfield  {author} {\bibinfo {author} {\bibfnamefont {H.}~\bibnamefont
  {Evertz}}, \bibinfo {author} {\bibfnamefont {H.}~\bibnamefont {Erkinger}}, \
  and\ \bibinfo {author} {\bibfnamefont {W.}~\bibnamefont {{Von der Linden}}},\
  }in\ \href@noop {} {\emph {\bibinfo {booktitle} {Computer Simulation Studies
  in Condensed Matter Physics XIV}}}\ (\bibinfo  {publisher} {Springer},\
  \bibinfo {year} {2002})\ pp.\ \bibinfo {pages} {123--123}\BibitemShut
  {NoStop}%
\bibitem [{\citenamefont {Grassberger}(1999)}]{Grassberger_1999}%
  \BibitemOpen
  \bibfield  {author} {\bibinfo {author} {\bibfnamefont {P.}~\bibnamefont
  {Grassberger}},\ }\href
  {https://www.sciencedirect.com/science/article/pii/S037843719800435X?via%3Dihub}
  {\bibfield  {journal} {\bibinfo  {journal} {Physica A}\ }\textbf {\bibinfo
  {volume} {262}},\ \bibinfo {pages} {251} (\bibinfo {year}
  {1999})}\BibitemShut {NoStop}%
\bibitem [{\citenamefont {Smirnov}(2001)}]{Smirnov_2001}%
  \BibitemOpen
  \bibfield  {author} {\bibinfo {author} {\bibfnamefont {S.}~\bibnamefont
  {Smirnov}},\ }\href {\doibase 10.1016/S0764-4442(01)01991-7} {\bibfield
  {journal} {\bibinfo  {journal} {C. R. Acad. Sci. Paris S{\' e}r. I Math.}\
  }\textbf {\bibinfo {volume} {333}},\ \bibinfo {pages} {239} (\bibinfo {year}
  {2001})}\BibitemShut {NoStop}%
\bibitem [{\citenamefont {Jacobsen}\ and\ \citenamefont
  {Zinn-Justin}(2002{\natexlab{a}})}]{Jacobsen_2002}%
  \BibitemOpen
  \bibfield  {author} {\bibinfo {author} {\bibfnamefont {J.~L.}\ \bibnamefont
  {Jacobsen}}\ and\ \bibinfo {author} {\bibfnamefont {P.}~\bibnamefont
  {Zinn-Justin}},\ }\href {\doibase 10.1088/0305-4470/35/9/304} {\bibfield
  {journal} {\bibinfo  {journal} {J. Phys. A: Math. Gen.}\ }\textbf {\bibinfo
  {volume} {35}},\ \bibinfo {pages} {2131} (\bibinfo {year}
  {2002}{\natexlab{a}})}\BibitemShut {NoStop}%
\bibitem [{\citenamefont {Jacobsen}\ and\ \citenamefont
  {Zinn-Justin}(2002{\natexlab{b}})}]{Jacobsen_2002_2}%
  \BibitemOpen
  \bibfield  {author} {\bibinfo {author} {\bibfnamefont {J.~L.}\ \bibnamefont
  {Jacobsen}}\ and\ \bibinfo {author} {\bibfnamefont {P.}~\bibnamefont
  {Zinn-Justin}},\ }\href {\doibase 10.1103/PhysRevE.66.055102} {\bibfield
  {journal} {\bibinfo  {journal} {Phys. Rev. E}\ }\textbf {\bibinfo {volume}
  {66}},\ \bibinfo {pages} {055102} (\bibinfo {year}
  {2002}{\natexlab{b}})}\BibitemShut {NoStop}%
\bibitem [{\citenamefont {Deng}\ \emph {et~al.}(2004)\citenamefont {Deng},
  \citenamefont {Bl\"ote},\ and\ \citenamefont {Nienhuis}}]{Deng_2004}%
  \BibitemOpen
  \bibfield  {author} {\bibinfo {author} {\bibfnamefont {Y.}~\bibnamefont
  {Deng}}, \bibinfo {author} {\bibfnamefont {H.~W.~J.}\ \bibnamefont
  {Bl\"ote}}, \ and\ \bibinfo {author} {\bibfnamefont {B.}~\bibnamefont
  {Nienhuis}},\ }\href {\doibase 10.1103/PhysRevE.69.026114} {\bibfield
  {journal} {\bibinfo  {journal} {Phys. Rev. E}\ }\textbf {\bibinfo {volume}
  {69}},\ \bibinfo {pages} {026114} (\bibinfo {year} {2004})}\BibitemShut
  {NoStop}%
\bibitem [{\citenamefont {Xu}\ \emph {et~al.}(2014)\citenamefont {Xu},
  \citenamefont {Wang}, \citenamefont {Zhou}, \citenamefont {Garoni},\ and\
  \citenamefont {Deng}}]{Xu_2014}%
  \BibitemOpen
  \bibfield  {author} {\bibinfo {author} {\bibfnamefont {X.}~\bibnamefont
  {Xu}}, \bibinfo {author} {\bibfnamefont {J.}~\bibnamefont {Wang}}, \bibinfo
  {author} {\bibfnamefont {Z.}~\bibnamefont {Zhou}}, \bibinfo {author}
  {\bibfnamefont {T.~M.}\ \bibnamefont {Garoni}}, \ and\ \bibinfo {author}
  {\bibfnamefont {Y.}~\bibnamefont {Deng}},\ }\href {\doibase
  10.1103/PhysRevE.89.012120} {\bibfield  {journal} {\bibinfo  {journal} {Phys.
  Rev. E}\ }\textbf {\bibinfo {volume} {89}},\ \bibinfo {pages} {012120}
  (\bibinfo {year} {2014})}\BibitemShut {NoStop}%
\bibitem [{\citenamefont {Elçi}\ \emph {et~al.}(2016)\citenamefont {Elçi},
  \citenamefont {Weigel},\ and\ \citenamefont {Fytas}}]{Elci_2016}%
  \BibitemOpen
  \bibfield  {author} {\bibinfo {author} {\bibfnamefont {E.~M.}\ \bibnamefont
  {Elçi}}, \bibinfo {author} {\bibfnamefont {M.}~\bibnamefont {Weigel}}, \
  and\ \bibinfo {author} {\bibfnamefont {N.~G.}\ \bibnamefont {Fytas}},\ }\href
  {\doibase https://doi.org/10.1016/j.nuclphysb.2015.12.001} {\bibfield
  {journal} {\bibinfo  {journal} {Nucl. Phys. B}\ }\textbf {\bibinfo {volume}
  {903}},\ \bibinfo {pages} {19 } (\bibinfo {year} {2016})}\BibitemShut
  {NoStop}%
\bibitem [{\citenamefont {Domb}(1974)}]{Domb_1974}%
  \BibitemOpen
  \bibfield  {author} {\bibinfo {author} {\bibfnamefont {C.}~\bibnamefont
  {Domb}},\ }\href {\doibase 10.1088/0305-4470/7/11/013} {\bibfield  {journal}
  {\bibinfo  {journal} {J. Phys. A}\ }\textbf {\bibinfo {volume} {7}},\
  \bibinfo {pages} {1335} (\bibinfo {year} {1974})}\BibitemShut {NoStop}%
\bibitem [{\citenamefont {Caracciolo}\ and\ \citenamefont
  {Sportiello}(2004)}]{caracciolo2004general}%
  \BibitemOpen
  \bibfield  {author} {\bibinfo {author} {\bibfnamefont {S.}~\bibnamefont
  {Caracciolo}}\ and\ \bibinfo {author} {\bibfnamefont {A.}~\bibnamefont
  {Sportiello}},\ }\href {\doibase 10.1088/0305-4470/37/30/002} {\bibfield
  {journal} {\bibinfo  {journal} {J. Phys. A}\ }\textbf {\bibinfo {volume}
  {37}},\ \bibinfo {pages} {7407} (\bibinfo {year} {2004})}\BibitemShut
  {NoStop}%
\bibitem [{\citenamefont {Tsallis}\ and\ \citenamefont
  {Levy}(1981)}]{Tsallis_1981}%
  \BibitemOpen
  \bibfield  {author} {\bibinfo {author} {\bibfnamefont {C.}~\bibnamefont
  {Tsallis}}\ and\ \bibinfo {author} {\bibfnamefont {S.~V.~F.}\ \bibnamefont
  {Levy}},\ }\href {\doibase 10.1103/PhysRevLett.47.950} {\bibfield  {journal}
  {\bibinfo  {journal} {Phys. Rev. Lett.}\ }\textbf {\bibinfo {volume} {47}},\
  \bibinfo {pages} {950} (\bibinfo {year} {1981})}\BibitemShut {NoStop}%
\bibitem [{\citenamefont {Stauffer}\ and\ \citenamefont
  {Aharony}(2018)}]{stauffer2018introduction}%
  \BibitemOpen
  \bibfield  {author} {\bibinfo {author} {\bibfnamefont {D.}~\bibnamefont
  {Stauffer}}\ and\ \bibinfo {author} {\bibfnamefont {A.}~\bibnamefont
  {Aharony}},\ }\href {\doibase 10.1201/9781315274386} {\emph {\bibinfo {title}
  {Introduction to percolation theory}}}\ (\bibinfo  {publisher} {CRC press},\
  \bibinfo {year} {2018})\BibitemShut {NoStop}%
\bibitem [{\citenamefont {Kesten}(1987)}]{kesten1987percolation}%
  \BibitemOpen
  \bibinfo {editor} {\bibfnamefont {H.}~\bibnamefont {Kesten}},\ ed.,\ \href
  {\doibase 10.1007/978-1-4613-8734-3} {\emph {\bibinfo {title} {Percolation
  theory and ergodic theory of infinite particle systems}}}\ (\bibinfo
  {publisher} {Springer-Verlag New York},\ \bibinfo {year} {1987})\BibitemShut
  {NoStop}%
\bibitem [{\citenamefont {Smirnov}\ and\ \citenamefont
  {Werner}(2001)}]{smirnov2001critical}%
  \BibitemOpen
  \bibfield  {author} {\bibinfo {author} {\bibfnamefont {S.}~\bibnamefont
  {Smirnov}}\ and\ \bibinfo {author} {\bibfnamefont {W.}~\bibnamefont
  {Werner}},\ }\href {https://dx.doi.org/10.4310/MRL.2001.v8.n6.a4} {\bibfield
  {journal} {\bibinfo  {journal} {Math. Res. Lett.}\ }\textbf {\bibinfo
  {volume} {8}},\ \bibinfo {pages} {729} (\bibinfo {year} {2001})}\BibitemShut
  {NoStop}%
\bibitem [{\citenamefont {Beffara}\ \emph {et~al.}(2011)\citenamefont
  {Beffara}, \citenamefont {Nolin} \emph {et~al.}}]{beffara2011monochromatic}%
  \BibitemOpen
  \bibfield  {author} {\bibinfo {author} {\bibfnamefont {V.}~\bibnamefont
  {Beffara}}, \bibinfo {author} {\bibfnamefont {P.}~\bibnamefont {Nolin}},
  \emph {et~al.},\ }\href {\doibase 10.1214/10-AOP581} {\bibfield  {journal}
  {\bibinfo  {journal} {Ann. Probab.}\ }\textbf {\bibinfo {volume} {39}},\
  \bibinfo {pages} {1286} (\bibinfo {year} {2011})}\BibitemShut {NoStop}%
\bibitem [{\citenamefont {Aizenman}\ \emph {et~al.}(1999)\citenamefont
  {Aizenman}, \citenamefont {Duplantier},\ and\ \citenamefont
  {Aharony}}]{aizenman1999path}%
  \BibitemOpen
  \bibfield  {author} {\bibinfo {author} {\bibfnamefont {M.}~\bibnamefont
  {Aizenman}}, \bibinfo {author} {\bibfnamefont {B.}~\bibnamefont
  {Duplantier}}, \ and\ \bibinfo {author} {\bibfnamefont {A.}~\bibnamefont
  {Aharony}},\ }\href {\doibase 10.1103/PhysRevLett.83.1359} {\bibfield
  {journal} {\bibinfo  {journal} {Phys. Rev. Lett.}\ }\textbf {\bibinfo
  {volume} {83}},\ \bibinfo {pages} {1359} (\bibinfo {year}
  {1999})}\BibitemShut {NoStop}%
\bibitem [{\citenamefont {Grossman}\ and\ \citenamefont
  {Aharony}(1987)}]{Grossman_1987}%
  \BibitemOpen
  \bibfield  {author} {\bibinfo {author} {\bibfnamefont {T.}~\bibnamefont
  {Grossman}}\ and\ \bibinfo {author} {\bibfnamefont {A.}~\bibnamefont
  {Aharony}},\ }\href {\doibase 10.1088/0305-4470/20/17/011} {\bibfield
  {journal} {\bibinfo  {journal} {J. Phys. A: Math. Gen.}\ }\textbf {\bibinfo
  {volume} {20}},\ \bibinfo {pages} {L1193} (\bibinfo {year}
  {1987})}\BibitemShut {NoStop}%
\bibitem [{\citenamefont {Saleur}\ and\ \citenamefont
  {Duplantier}(1987)}]{Saleur_1987}%
  \BibitemOpen
  \bibfield  {author} {\bibinfo {author} {\bibfnamefont {H.}~\bibnamefont
  {Saleur}}\ and\ \bibinfo {author} {\bibfnamefont {B.}~\bibnamefont
  {Duplantier}},\ }\href {\doibase 10.1103/PhysRevLett.58.2325} {\bibfield
  {journal} {\bibinfo  {journal} {Phys. Rev. Lett.}\ }\textbf {\bibinfo
  {volume} {58}},\ \bibinfo {pages} {2325} (\bibinfo {year}
  {1987})}\BibitemShut {NoStop}%
\bibitem [{\citenamefont {Coniglio}(1989)}]{Coniglio_1989}%
  \BibitemOpen
  \bibfield  {author} {\bibinfo {author} {\bibfnamefont {A.}~\bibnamefont
  {Coniglio}},\ }\href {\doibase 10.1103/PhysRevLett.62.3054} {\bibfield
  {journal} {\bibinfo  {journal} {Phys. Rev. Lett.}\ }\textbf {\bibinfo
  {volume} {62}},\ \bibinfo {pages} {3054} (\bibinfo {year}
  {1989})}\BibitemShut {NoStop}%
\bibitem [{\citenamefont {Huber}()}]{Huber}%
  \BibitemOpen
  \bibfield  {author} {\bibinfo {author} {\bibfnamefont {G.}~\bibnamefont
  {Huber}},\ }\href@noop {} {\bibinfo  {journal} {unpublished}\ }\BibitemShut
  {NoStop}%
\bibitem [{\citenamefont {Ferrenberg}\ \emph {et~al.}(2018)\citenamefont
  {Ferrenberg}, \citenamefont {Xu},\ and\ \citenamefont
  {Landau}}]{Ferrenberg2018Ising}%
  \BibitemOpen
\bibfield  {journal} {  }\bibfield  {author} {\bibinfo {author} {\bibfnamefont
  {A.~M.}\ \bibnamefont {Ferrenberg}}, \bibinfo {author} {\bibfnamefont
  {J.}~\bibnamefont {Xu}}, \ and\ \bibinfo {author} {\bibfnamefont {D.~P.}\
  \bibnamefont {Landau}},\ }\href {\doibase 10.1103/PhysRevE.97.043301}
  {\bibfield  {journal} {\bibinfo  {journal} {Phys. Rev. E}\ }\textbf {\bibinfo
  {volume} {97}},\ \bibinfo {pages} {043301} (\bibinfo {year}
  {2018})}\BibitemShut {NoStop}%
\bibitem [{\citenamefont {Hou}\ \emph {et~al.}(2019)\citenamefont {Hou},
  \citenamefont {Fang}, \citenamefont {Wang}, \citenamefont {Hu},\ and\
  \citenamefont {Deng}}]{Hou2019FKIsing}%
  \BibitemOpen
  \bibfield  {author} {\bibinfo {author} {\bibfnamefont {P.}~\bibnamefont
  {Hou}}, \bibinfo {author} {\bibfnamefont {S.}~\bibnamefont {Fang}}, \bibinfo
  {author} {\bibfnamefont {J.}~\bibnamefont {Wang}}, \bibinfo {author}
  {\bibfnamefont {H.}~\bibnamefont {Hu}}, \ and\ \bibinfo {author}
  {\bibfnamefont {Y.}~\bibnamefont {Deng}},\ }\href {\doibase
  10.1103/PhysRevE.99.042150} {\bibfield  {journal} {\bibinfo  {journal} {Phys.
  Rev. E}\ }\textbf {\bibinfo {volume} {99}},\ \bibinfo {pages} {042150}
  (\bibinfo {year} {2019})}\BibitemShut {NoStop}%
\bibitem [{\citenamefont {Lundow}\ and\ \citenamefont
  {Markstr{\"o}m}(2009)}]{lundow2009critical}%
  \BibitemOpen
  \bibfield  {author} {\bibinfo {author} {\bibfnamefont {P.~H.}\ \bibnamefont
  {Lundow}}\ and\ \bibinfo {author} {\bibfnamefont {K.}~\bibnamefont
  {Markstr{\"o}m}},\ }\href {\doibase 10.1103/PhysRevE.80.031104} {\bibfield
  {journal} {\bibinfo  {journal} {Phys. Rev. E}\ }\textbf {\bibinfo {volume}
  {80}},\ \bibinfo {pages} {031104} (\bibinfo {year} {2009})}\BibitemShut
  {NoStop}%
\bibitem [{\citenamefont {Lundow}\ and\ \citenamefont
  {Markstr{\"o}m}(2014)}]{lundow2014finite}%
  \BibitemOpen
  \bibfield  {author} {\bibinfo {author} {\bibfnamefont {P.~H.}\ \bibnamefont
  {Lundow}}\ and\ \bibinfo {author} {\bibfnamefont {K.}~\bibnamefont
  {Markstr{\"o}m}},\ }\href {\doibase 10.1016/j.nuclphysb.2014.10.011}
  {\bibfield  {journal} {\bibinfo  {journal} {Nucl. Phys. B}\ }\textbf
  {\bibinfo {volume} {889}},\ \bibinfo {pages} {249} (\bibinfo {year}
  {2014})}\BibitemShut {NoStop}%
\bibitem [{\citenamefont {Madras}\ and\ \citenamefont
  {Sokal}(1988)}]{madras1988pivot}%
  \BibitemOpen
  \bibfield  {author} {\bibinfo {author} {\bibfnamefont {N.}~\bibnamefont
  {Madras}}\ and\ \bibinfo {author} {\bibfnamefont {A.~D.}\ \bibnamefont
  {Sokal}},\ }\href {\doibase 10.1007/BF01022990} {\bibfield  {journal}
  {\bibinfo  {journal} {J. Stat. Phys.}\ }\textbf {\bibinfo {volume} {50}},\
  \bibinfo {pages} {109} (\bibinfo {year} {1988})}\BibitemShut {NoStop}%
\end{thebibliography}%

\onecolumngrid
\clearpage

\begin{center}

{\large \bf Supplemental Material for\\[0.2em]
``Loop-Cluster Coupling and Algorithm for Classical Statistical Models''}\\[1em]

Lei Zhang${}^{1,2}$, Manon Michel${}^{3}$, Eren M. El{\c{c}}i${}^{4}$, and Youjin Deng${}^{1,2,5}$\\[0.5em]
{\small ${}^1$\it Hefei National Laboratory for Physical Sciences at Microscale and Department of Modern Physics, University of Science and Technology of China, Hefei, Anhui 230026, China\\[-0.05em]
${}^2$CAS Center for Excellence and Synergetic Innovation Center in Quantum Information and Quantum Physics, University of Science and Technology of China, Hefei, Anhui 230026, China \\[-0.05em]
${}^3$CNRS, Laboratoire de math\'ematiques Blaise Pascal, UMR 6620, Universit\'e Clermont-Auvergne, Aubi\`ere, France  \\[-0.05em]
${}^4$ School of Mathematical Sciences, Monash University, Clayton, VIC 3800, Australia  \\[-0.05em]
${}^5$ Department of Physics and Electronic Information Engineering, Minjiang University, Fuzhou, Fujian 350108, China \\[-0.05em]
}

\end{center}

\vspace{0.3cm}

\twocolumngrid

\section{Loop-cluster algorithm for real $q$ and a single-cluster version}

The loop-cluster (LC) algorithm can be extended to sample from
the RC model of real value $q \geq 1$,
via the induced-subgraph decomposition~\cite{deng2007cluster}.
Starting with a FK bond configuration and setting an integer
$1 \leq m \leq q$, each cluster is randomly picked as ``active" with probability $m/q$ or sampled as ``inactive". 
 One obtains then an
 effective RC model with $q'=m$ on the subgraph defined by active vertices and edges 
  and a model with $q'=q-m$ on the complementary inactive subgraph.  
  The active partition can then be updated through
  any valid MC algorithms, while the inactive one is left unchanged, which is effectively an identity operation.
  For $ 2 \! >\! q \geq 1$, with the unique choice $m \! = \! 1$, one
can apply the conventional percolation strategy for any active
  edge, corresponding to the Chayes-Machta
algorithm~\cite{chayes1998graphical}.  For $q \geq 2$, one can choose
integer $m \geq 2$ and apply the LC algorithm on the active
 subgraph, leading to an extended LC algorithm.

Moreover, a single-cluster version
can be formulated to sample from the $q$-flow model.
Starting from a $q$-flow configuration,
one randomly chooses a root vertex and grows a cluster by Step A
until it cannot become larger, i.e. all the boundary edges have been
sampled as unoccupied; then a new $q$-flow configuration can be
sampled through Step B.  Like the Wolff algorithm, the single-cluster LC
algorithm is more likely to update larger clusters~\cite{wolff1989collective}, which on average
contain larger loops, and to show higher efficiency.

\section{Numerical study of dynamical behavior}

We study the dynamics of the LC algorithm
and compare it to the Swendsen-Wang (SW) scheme for both ``energy-'' and
``susceptibility-like" quantities in the FK representation,
i.e. respectively the total number $\mathcal{N}$ of occupied bonds and
the second moment of FK cluster sizes, defined as
$\mathcal{S}_2 =  \sum_C|C|^2$ with $|C|$ the size of cluster $C$.

In the LC scheme, the number of operations in Step A equals the number of edges of the
original graph, $|E|$, and is $2|E|$ in Step B.
The total number of operations is $3|E|$, slightly larger than $2|E|$ for the SW algorithm.

Simulations are performed on toroidal grids for
$2  \leq   d   \leq  5$ and on finite complete graphs (CG)
with $n$ vertices.  The critical coupling strengths are
$J_{\rm c} \! =\! \ln(\sqrt{q}\! +\! 1)$ for
$(q\! =\!2, 3, d \! = \! 2)$, $ 0.443 \, 309 \, 262 (16)$ for
$(q\! =\!2, d \! = \! 3)$~\cite{Ferrenberg2018Ising,Hou2019FKIsing},
$0.299\, 389 \, 4(10) $ for
$(q\! =\!2, d \! = \! 4)$~\cite{lundow2009critical},
$0.227 \, 830 \, 0 (8)$ for
$(q\! =\!2, d \! = \! 5)$~\cite{lundow2014finite}, and $2/n$ for
$(q\! =\!2, {\rm CG})$.

For an observable $\mathcal{O}$, we calculate the normalized
autocorrelation function
\begin{equation}
\rho_{\mathcal{O}}(t) \!=\! (\langle\mathcal{O}_0\mathcal{O}_t\rangle
\!-\! \langle\mathcal{O}\rangle^2)/(\langle\mathcal{O}^2\rangle \!-\!
\langle\mathcal{O}\rangle^2)
\end{equation}
and the integrated autocorrelation time
\begin{equation}
   \tau_{\text{int},\mathcal{O}} \!=\!  \frac{1}{2} \!+\!
   \sum_{t=1}^{+\infty} \rho_{\mathcal{O}}(t),
 \end{equation}
 where the time unit corresponds to a configuration update.  
 In practice, we use the windowing method~\cite{madras1988pivot}
 to truncate the summation for $\tau_{\rm int}$. In our data analysis,
 the windowing parameter $c$ was chosen to be 6 for ($d = 2, q = 3$) and to be 8
 otherwise to give good estimates. 
 
  \begin{table}
\begin{tabular}{|c|lllll|c|}
\hline
$(q,d)$                 & $(2,2{\rm d})$   & $(2,3{\rm d})$     & $(2,4{\rm d})$    & $(2,5{\rm d})$    &$(2,{\rm CG})$     &$(3,2{\rm d})$     \\
 \hline
$\mathcal{N}$      & 0.16(1)    & 0.48(2)     & 0.62(2)       & 0.92(9)      & 0.215(6)     & 0.48(2) \\
\hline
$\mathcal{S}_2$  & 0.17(1)     & 0.40(1)     & 0.69(1)      & 0.99(2)      & 0.259(2)     & 0.45(1) \\
\hline
\end{tabular}
\caption{Dynamical critical exponents $z_{\text{int}}$ for the LC algorithm obtained for different values ($q, d$) and observables.}
\label{tab:z}
\end{table}
 
 We have $5\times 10^7$ to $10^8$ samples for each $(q,d,L)$ or $(q,n)$, 
 where $L$ is the linear system size and the number of vertices for the complete graph is set as $n=L^2$.
 According to the least-squares criterion, the $\tau_{\rm int} (L)$ data are fitted by $A+BL^{z_{\rm int}}$,
 where $z_{\rm int}$ is the dynamical critical exponent and $A$ and $B$ are non-universal constants.
 In practice, we gradually increase the smallest system size $L_{\rm min}$ such that 
 the data for $ L < L_{\text{min}}$ are excluded from the fit until the ratio of $\chi^2$ and degree of freedom (DF)
 is close to 1 and subsequent increases of $L_{\text{min}}$  do not cause the $\chi^2$ value to drop by vastly
more than one unit per degree of freedom. The estimates of $z_{\rm int}$ are given in Table~\ref{tab:z},
which are consistent with the results for the SW algorithm reported in the literature. 

 \begin{figure*}[t]
% \centering
\includegraphics[width=1.0\textwidth]{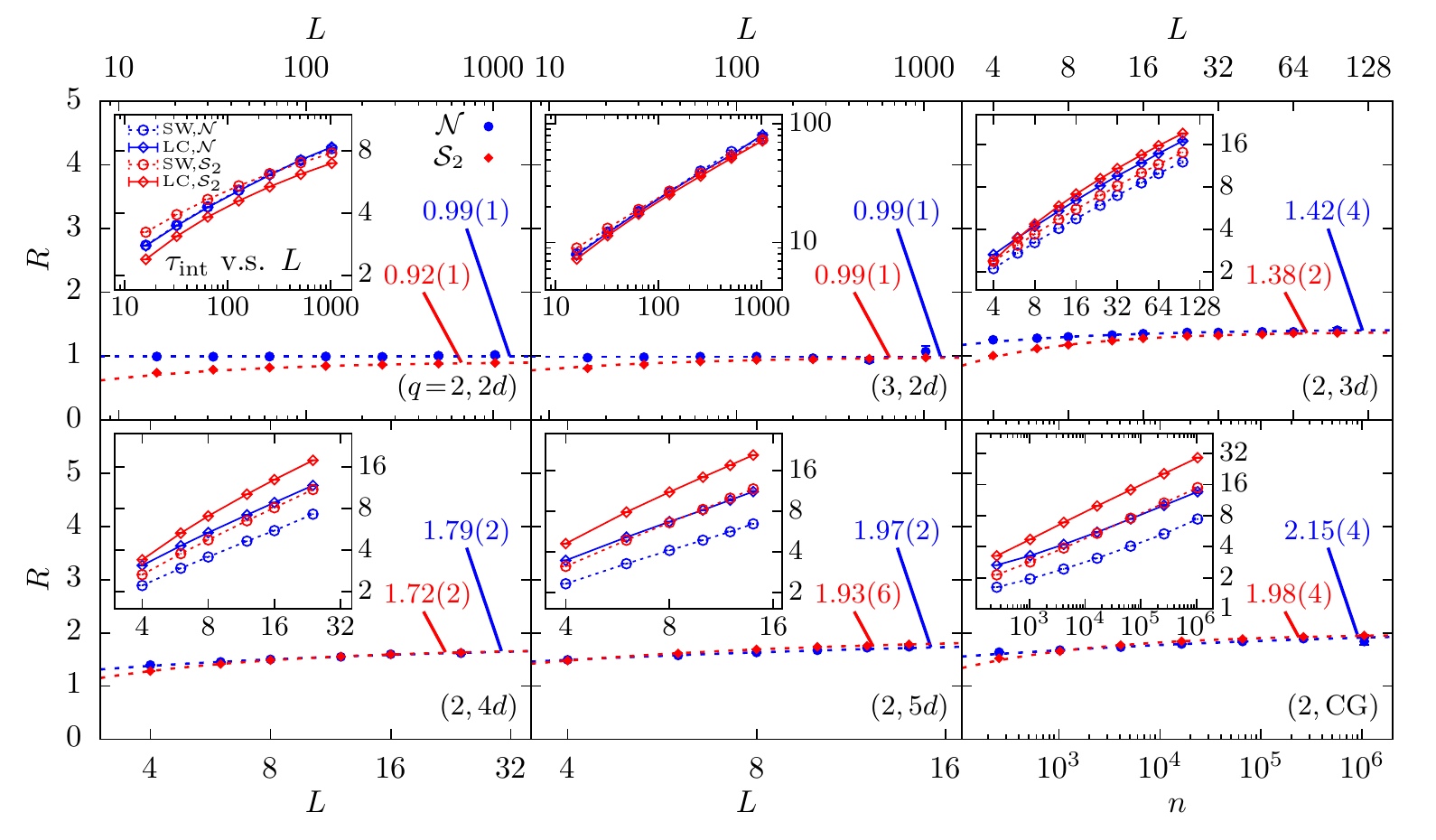}
\caption{Ratios of integrated autocorrelation times
  $R \! = \! \tau_{\rm int, LC}/ \tau_{\rm int,SW}$ for the LC and the
  SW algorithm, with $q \! = \! 2$ in dimensions
  $ 2 \! \leq \!  d \! \leq \! 5$ and on the complete graph (CG), as
  well as with $(q \! = \! 3, d \! = \! 2)$.  The values of
  $ \tau_{\rm int}$ are shown in the inset plots and the
    asymptotic fitted values for $R_{\text{inf}}$ are indicated in each subplot. }
\label{fig:tauS2}
\end{figure*}
 
 To further check whether the LC and the SW algorithm are in the same dynamical universality class, 
 we measure the integrated correlation time $\tau_{\rm int}$ for the SW scheme and
 calculate the ratio  $R \!=\! \tau_{\rm int, LC}/\tau_{\rm int, SW}$. 
 The results of $R$ are shown in Fig.~\ref{fig:tauS2}, where the insets display the
 $\tau_{\rm int}$ values for both the LC and SW methods.
 It is clear that for both energy- and susceptibility-like quantities
 $\mathcal{N}$ and $\mathcal{S}_2$, the ratio $R$ converges to a
 constant as system size increases. In two dimensions, it is
 interesting to observe that $R_{\mathcal{N}}$ is consistent with $1$,
 irrespective of system size $L$ and the $q$ value. 
 For each $(q,d)$, the $R (L)$ data are fitted by ansatz
 $A+BL^{-\Delta}$, with $\Delta$ a correction exponent. The
 fitting results are shown in Table \ref{tab:A0} and the asymptotic values of
 $A$ are also displayed in Fig.~\ref{fig:tauS2}. It
 exhibits an increase of the $A$ value with $d$, up to a value
 slightly larger than 2 for CG (effectively
 $d \! \rightarrow \! \infty$).  Therefore, it is strongly suggested
 that the LC and the SW algorithm belong to the same dynamical
 universality class.

\begin{table}
\begin{tabular}{|ll|lll|lc|}
\hline
 Obs      &$(q,d)$                  & Fit                           &$A$          &   $\Delta$ & $L_{\text{min}}$   & $\chi^2/\text{DF}$\\
\hline
  $\mathcal{N}$     & $(2, 2d)$            & $A$                         & 0.99(1)  \;\;     & ---            & 32           & 3.1/5\\
                              & $(2, 3d)$            & $A+BL^{-\Delta}$    & 1.42(4)       &0.7(2)      & 8             & 6.2/5\\
                              & $(2, 4d)$            &  $A+BL^{-\Delta}$    & 1.79(2)    &0.5          & 6             & 3.8/3\\
                              & $(2, 5d)$            & $A+BL^{-\Delta}$     & 1.97(2)    &0.5         & 4               & 4.5/4\\
                              & $(2, {\rm CG})$  & $A+BL^{-\Delta}$       & 2.15(4)    & 0.2      & 32         & 1.1/3 \\
 \hline
  $\mathcal{S}_2$  & $(2, 2d)$           & $A+BL^{-\Delta}\;\;$    & 0.92(1)    &0.5           & 128         & 1.6/2\\
                              & $(2, 3d)$            & $A+BL^{-\Delta}$    & 1.38(2)       &1.0(1)      & 8             & 6.3/5\\
                              & $(2, 4d)$            & $A+BL^{-\Delta}$    & 1.72(2)    &0.9(1)      & 4             & 2.7/3\\
                              & $(2, 5d)$            &  $A+BL^{-\Delta}$     & 1.93(6)    &0.9(1)    & 4             & 3.1/3\\
                              & $(2, {\rm CG})$  &  $A+BL^{-\Delta}$      & 1.98(4)    &0.6(2)   & 32        & 2.4/3\\
\hline
$\mathcal{N}$      & $(3, 2d)$             &  $A$                             & 0.99(1)   &---         & 64               & 5.6/3\\
$\mathcal{S}_2$   &                            & $A+BL^{-\Delta}$       & 0.99(1)    &0.5      & 64              & 4.0/3\\
\hline
\end{tabular}
   \caption{Fitting results for
     $R \! = \! \tau_{\rm int, LC}/ \tau_{\rm int,SW}$
     as a function of $q$, $d$ and observables (Obs), by
     ansatz $A+BL^{-\Delta}$, with $\Delta$ a correction
     exponent. ``DF" means degree of freedom.
   }
   \label{tab:A0}
\vspace{-5mm}
\end{table}

\begin{figure*}[t]
    \centering
\includegraphics[width=1.0\textwidth]{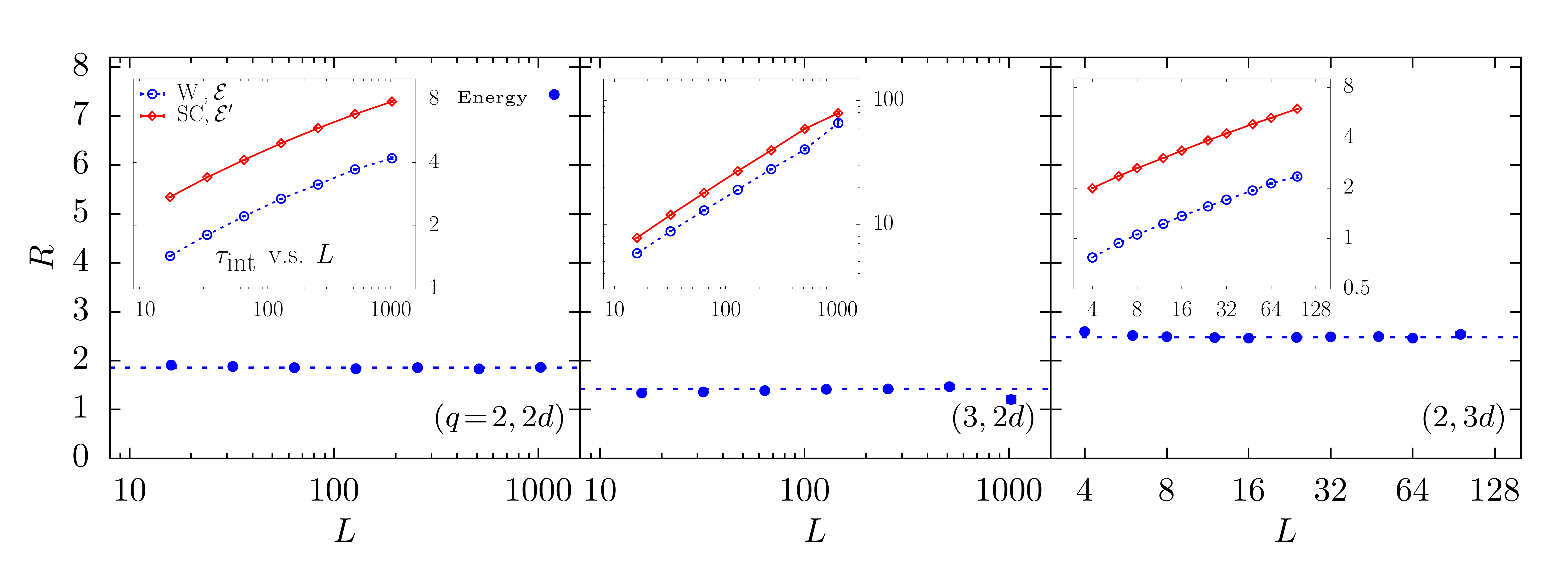}
\caption{Ratios of integrated autocorrelation times
  $R \! = \! \tau_{\rm int, LC}/ \tau_{\rm int,W}$ for the LC
  single-cluster variant and the Wolff algorithm, with $q \! = \! 2$
  in dimensions $2$ and $3$ for the energy, as well as with
  $q \! = \! 3$ with $d \! = \! 2$. The asymptotic fitted values for $A$ are $1.85(2)$, $2.48(2)$ and $1.42(2)$ for $(q=2, 2d)$, $(2,3d)$ and $(3,2d)$,
  respectively. The values of $ \tau_{\rm int}$ are shown in the insets. }
\label{fig:tauWolff}
\end{figure*}

  We also compare the dynamical behavior of the single-cluster LC algorithm and the Wolff method.
  Similarily, we measure the integrated correlation time $\tau_{\rm int}$,
  and calculate the ratio $R = \tau_{\rm int, LC}/ \tau_{\rm int,W}$.
  Our numerical results confirm that the single-cluster LC algorithm
  and the Wolff method have the same average sizes of the updated
  cluster and belong to the same dynamical class, as illustrated by
  Fig.~\ref{fig:tauWolff} where the scaling of the corresponding
  integrated autocorrelation times is displayed. The simulations were
  carried out on a hypercubic lattice, for $(q\! = \! 2, d \!=\!2,3)$ (windowing parameter
  set to $8$) and $(q\!=\!3, d\!=\!2)$ (windowing parameter set to $6$). In
  total, $10^8$ measurements of an ``energy-like'' quantity (number of
  edges $\mathcal{E}$ connecting spins of the same values for the Wolff algorithm in the FK representation and
  number of non-trivial flows $\mathcal{E}'$ in the $q$-flow representation for the
  LC single-cluster variant) were made. The time unit is normalized to
  one full system size sweep for consistency.

\end{document}